\documentclass[aps,prd,twocolumn,letterpaper,showpacs,floatfix]{revtex4}
\usepackage[squaren]{SIunits}
\usepackage{amssymb,amsmath}
\usepackage{graphicx}
\usepackage{dcolumn}
\newcolumntype{.}{D{.}{.}{-1}}
\usepackage{hyperref}

\begin{document}

\newcommand{\Eref}[1]{Eq.~(\ref{#1})}
\newcommand{\Erefs}[2]{Eq.~\eqref{#1}--\eqref{#2}}

\newcommand{\Fig}[1]{Fig.~\ref{#1}}
\newcommand{\Figure}[1]{Figure~\ref{#1}}
\newcommand{\Sec}[1]{Sec.~\ref{#1}}
\newcommand{\Tab}[1]{Tab.~\ref{#1}}

\newcommand{\Pizza}{\textsc{pizza\ }}

\newcommand{\D}{\partial}
\newcommand{\Dc}{\nabla}
\newcommand{\Rmd}{\rho}
\newcommand{\Rmdp}{\rho_p}
\newcommand{\Sed}{\epsilon}
\newcommand{\Sent}{h}
\newcommand{\Press}{P}
\newcommand{\Csnd}{c_s}
\newcommand{\Crmd}{D}
\newcommand{\Ced}{\tau}
\newcommand{\Cmom}{S}
\newcommand{\Sam}{l_\varphi}
\newcommand{\Ped}{U}
\newcommand{\Pam}{L}
\newcommand{\Amp}{A}
\newcommand{\Sg}{\sqrt{g}}
\newcommand{\Vc}{\sqrt{\gamma}}
\newcommand{\Lf}{W}
\newcommand{\Lapse}{\alpha}
\newcommand{\Shift}{\beta}
\newcommand{\Gpp}{g_{\phi\phi}}
\newcommand{\Half}{\frac{1}{2}}
\newcommand{\Inv}[1]{\frac{1}{#1}}
\newcommand{\e}[1]{\ensuremath{\cdot 10^{#1}}}
\newcommand{\parsec}{\ensuremath{\mathrm{pc}}}
\newcommand{\BUthree}{MB85}
\newcommand{\BUsix}{MB70}

\title{Nonlinear Decay of $r$ modes in Rapidly Rotating Neutron Stars}
\author{Wolfgang Kastaun}
\affiliation{International School for Advanced Studies (SISSA), via Bonomea 265, Trieste 34136, Italy}
\affiliation{Istituto Nazionale di Fisica Nucleare (INFN), Via Enrico Fermi 40, 00044 Frascati (Rome), Italy}
\date{\today}

\begin{abstract}
We investigate the dynamics of $r$ modes
at amplitudes in the nonlinear regime
for rapidly but uniformly rotating neutron stars with a polytropic equation of state. 
For this, we perform three-dimensional relativistic 
hydrodynamical simulations, 
making the simplifying assumption of a fixed spacetime.
We find that for initial dimensionless amplitudes around three,
$r$ modes decay on timescales around ten oscillation periods,
while at amplitudes of order unity, they are stable over the evolution timescale.
Together with the decay, a strong differential rotation develops,
conserving the total angular momentum, 
with angular velocities in the range $0.5\dots1.2$ of the initial one.
By comparing two models, we found that increasing rotation slows down the 
$r$-mode decay.
We present $r$-mode eigenfunctions and frequencies,
and compare them to known analytic results for slowly rotating Newtonian stars.
As a diagnostic tool, 
we discuss conserved energy and angular momentum 
for the case of a fixed axisymmetric background metric and
introduce a measure for the energy of non-axisymmetric fluid oscillation modes.
\end{abstract}

\maketitle
\section{Introduction}

The $r$ modes of neutron stars are a subclass of inertial modes, i.e. modes
where the Coriolis force is the main restoring force, 
which have purely axial parity in the slow rotation limit.
In the limit of slowly and rigidly rotating  Newtonian stars, 
the $r$-mode eigenfunction and frequency is analytically known.
Its frequency, as for all inertial modes, is proportional to the
angular velocity $\Omega$ of the star.
The $r$ modes are current dominated, that is the amplitude of the density perturbations
is smaller than the velocity perturbations by a factor $\Omega$.
Because of this and also due to the lower frequencies, 
they are weak emitters of gravitational waves compared
to modes of same energy for which pressure is the dominant restoring force.

The extension of the $r$-mode solution to the case of rapid rotation and/or
General Relativity (GR) is an ongoing field of research.
In the slow rotation approximation, GR equations have first been derived by
\cite{Kojima92, Kojima98}. 
In~\cite{Lockitch2000} it was shown that they are valid only for 
non-barotropic equations of state (EOS), 
and that the equations for barotropic stars are qualitatively different.
Further corrections to the equations were pointed out by~\cite{Ruoff2002}.
The equations in~\cite{Kojima98,Ruoff2002} for non-barotropic stars contain singular points,
whose interpretation is still under debate 
\cite{Lockitch2004, Lockitch03, Lockitch2000, Lockitch99, Beyer99, Yoshida2001, Yoshida2002, Ruoff2001, Kokkotas03}.
In particular, \cite{Lockitch2004} provides arguments that the slow rotation approximation 
breaks down near the singular points, and including higher order terms would lead to 
valid physical solutions. 
However, there is still no mathematical proof that regular $r$ modes of 
non-barotropic stars in GR exist under all circumstances.
For the barotropic case, i.e. an EOS where the pressure is a function of density alone,
GR counterparts of the Newtonian $r$ mode have been found by~\cite{Lockitch03}.
Those solutions are hybrid modes containing both polar and axial parity perturbations.

The rapidly rigidly rotating case has been investigated 
in GR only with the simplifying
assumption of a fixed spacetime (relativistic Cowling approximation).
In~\cite{Yoshida05}, the two-dimensional partial differential equations
governing the eigenfunctions for the case of a barotropic EOS
have been solved numerically by using finite differences.
It was found that the ratio of oscillation and rotation frequencies 
is decreasing with increasing ratio of rotational to binding energy.
In~\cite{Gaertig2008}, 
the $r$ modes and pressure modes of rapidly but rigidly rotating polytropic stars are
investigated using time-evolution of the linearized equations for a prescribed 
$\phi$-dependency.
The method is extended to include the non-barotropic case in \cite{Gaertig2009},
and differential rotation in \cite{Gaertig2010}.

The $r$ mode is of astrophysical interest 
since in GR it is generally subject to unstable growth, 
emitting gravitational radiation, 
as shown by~\cite{Andersson98, Friedmann98}.
The reason is that the $r$-mode wave pattern is generally 
counter-rotating in the co-rotating frame,
but co-rotating in the inertial frame, 
which is the condition for the 
CFS mechanism discovered by~\cite{Chandrasekhar70, Friedman78} to be operational.
Due to the CFS instability, 
$r$ modes are a possible source for gravitational wave astronomy,
but might also explain the limitation of observed neutron star rotation rates.
It is thus important to know at which amplitudes the $r$-mode instability saturates.
There are several effects which might suppress the instability or limit the amplitude to small values,
including nonlinear couplings~\cite{Arras2003, Brink2004, Lin2006}, 
winding up of magnetic field lines by differential rotation associated with the $r$ mode 
\cite{Rezzolla2000,Rezzolla2001, Rezzolla2001b},
and bulk viscosity~\cite{Lindblom2002} enhanced by the presence of hyperons.
The studies mentioned so far apply to newborn neutron stars, 
which are still hot enough to prevent superfluidity,
which further complicates the picture, see~\cite{Lee2003, Glampedakis2006, Mendell2000}.

Assuming the $r$ mode has an instability window, 
i.e. a range of temperature and
rotation rate where the CFS instability is not suppressed,
it would be important to know the behavior at very high amplitudes,
typically expressed as a dimensionless quantity proportional to the ratio
of velocity perturbation to rotational velocity at the equator.
Using numerical simulations of uniformly rotating stars in the Newtonian framework, 
it was found by~\cite{Tohline2002} that formation of shocks and breaking of surface waves set in at 
dimensionless amplitudes around three.
However, the $r$ modes in these simulations have been continuously excited by a
gravitational back-reaction force artificially increased to huge values.
In a similar simulation~\cite{Gressman2002,Lin2006}, 
but switching off the artificial driving force after amplitudes
in the range $1.6\dots 2.2$ had been reached, 
a catastrophic decay of the $r$ modes was found.
This decay was attributed to mode coupling effects, 
due to the observed growth of secondary modes.
In full GR, 
a numerical simulation of an $r$ mode with unit amplitude
was performed  by~\cite{Font2001},
but no decay apart from numerical damping was found on the evolution timescale
of 26 oscillation periods.

The $r$ modes are also linked to differential rotation in many ways.
For the case of rapidly and differentially rotating barotropic 
Newtonian stars,
\cite{Karino2001} found that the $r$-mode eigenvalue problem for a polytropic Newtonian star 
becomes difficult to solve numerically for degrees of differential rotation large enough to cause
corotation points of the wave pattern.
On the other hand, \cite{Yoshida2001b} found regular solutions for a Newtonian, 
incompressible thin shell model with a similar amount of differential rotation.
The existence of regular $r$-mode solutions for differentially rotating stars is thus likely, 
but remains unproven.
For uniformly rotating stars, the presence of $r$ modes implies a certain amount of differential
rotation, which is of second order in the mode amplitude.
In~\cite{Rezzolla2000,Rezzolla2001}, this effect was estimated using only the first order eigenfunctions.
The importance for the evolution of a magnetic field via winding up of field lines
was demonstrated in \cite{Rezzolla2000, Rezzolla2001b}.
Complementary to this, it was shown by~\cite{Sa2004} that the second order 
corrections to the eigenfunction
of the velocity field necessarily contain a differentially rotating part,
at least for slowly and rigidly rotating Newtonian stars with barotropic EOS.
As argued in~\cite{Sa2005, Sa2006}, the resulting second order contributions
to the angular momentum have to be considered 
for the time evolution of the $r$-mode instability.
A similar, but distinct effect concerns the interaction with gravitational radiation.
For a toy model consisting of a spherical shell, 
it was shown by~\cite{Levin2001} that the gravitational back-reaction 
directly induces differential rotation
instead of causing only a uniform spindown.
Further, the aforementioned studies~\cite{Tohline2002,Gressman2002} 
of nonlinear $r$-mode decay 
also observed the formation of significant differential rotation during the decay. 
These studies support the view that the CFS instability of the $r$ mode
will cause a certain amount of differential rotation, 
even if the $r$-mode amplitude saturates at small values.

In the present work, we extent the previous studies in mainly two directions.
First, we extract $r$-mode eigenfunctions and frequencies for rapidly (but rigidly) 
rotating relativistic stars under the simplifying Cowling approximation.
We are using a fundamentally different approach than~\cite{Yoshida05,Gaertig2008},
and also discuss the properties of the eigenfunctions in detail, 
in particular their energy and estimated gravitational radiation.
Second, we investigate the decay of high amplitude $r$ modes
and the formation of differential rotation already found in~\cite{Tohline2002, Gressman2002}.
We are however using the relativistic Cowling approximation 
instead of Newtonian gravity for the evolution
and the exact (linear) $r$-mode eigenfunctions as initial perturbation,
instead of breeding $r$ modes by using artificial back-reaction forces.
This will shed light on the robustness of those effects.

\section{Analytic tools}
In this section, we review the analytic tools used for this work.
Throughout the article, we use the following notation:
\begin{align*}
\Rmd    &\equiv \text{Rest frame rest mass density}\\
P       &\equiv \text{Pressure}\\
\Sed    &\equiv \text{Specific internal energy}\\
\Sent   &\equiv 1 + \Sed + \frac{P}{\Rmd} \\
u^\mu   &\equiv \text{Fluid 4-velocity}\\
v^i     &\equiv \text{Fluid 3-velocity}\\
\Lf     &\equiv \text{Fluid Lorentz factor}\\
g_{\mu\nu} &\equiv \text{4-metric of signature} (-,+,+,+)\\
g_{ij}  &\equiv \text{3-metric}\\
\Lapse  &\equiv \text{Lapse function}\\
\Shift^i &\equiv \text{Shift vector}\\
\Omega &\equiv \text{Angular velocity of the star}
\end{align*}
All equations assume geometric units $G=c=1$. 
Greek indices run from $0 \dots 3$, indices $i,j,k,l$ from $1 \dots 3$.
When working in cylindrical coordinates, generally denoted by $(\vartheta,z,\phi)$, 
indices $a,b,c$ run from $1 \dots 2$, excluding $\phi$.

\subsection{General properties of eigenfunctions}
Although there is no analytic solution for the oscillation eigenfunctions
of rapidly rotating stars, one can derive some general properties.
The following is valid for a fixed spacetime, 
rigid rotation, and a barotropic EOS.

For any global oscillation mode,
the perturbation of a fluid quantity $X$ can be written as
\begin{align}
\delta X\left(\vec{x},t\right) 
  &= \Amp \Re\left( \hat{X}\left(\vartheta,z\right)
     e^{i\left(\omega t+m\phi + \theta_X\right)}\right),  \label{eq_generic_mode}
\end{align}
where $\Amp$ is a dimensionless amplitude, 
$\omega$ the real-valued oscillation frequency,
and $\theta_X$ is a constant phase shift.
$\hat{X}$ is the (suitably normalized) real-valued eigenfunction.

The set of variables we use to completely specify the oscillation 
is $\{\Sed, v^i\}$.
The relative phase shifts are given by
$\theta_\Sed = \theta_{v^\phi} = 0$, $\theta_{v^\vartheta} = \theta_{v^z}=\frac{\pi}{2}$.
For a derivation, see~\cite{Kastaun08}.

Due to the equatorial symmetry of the unperturbed models,
the eigenfunctions also have a well defined z-parity, with the relations
\begin{align}
P_z[\delta\Sed] = P_z[\delta v^\vartheta] = -P_z[\delta v^z] = P_z[\delta v^\phi].
\end{align}
\subsection{Newtonian $r$ mode}
For slowly rotating stars in Newtonian theory,  
there exist analytic solutions for the $r$ modes.
The velocity eigenfunctions are given by 
\begin{align}\label{eq_ef_r_newt}
  \delta \vec{v} &= \Amp \Omega R \left(\frac{r}{R}\right)^l \Re\left(\vec{Y}^B_{ll} \right), 
\end{align}
where 
\begin{align}
  \vec{Y}^B_{lm} &= \left(l\left(l+1\right)\right)^{-\Half} \vec{r} \times \vec{\Dc} Y_{lm} 
\end{align}
are the pure-spin vector harmonics of magnetic type (see e.g.~\cite{Thorne80}).
The density perturbation is proportional to $\Omega^2 Y_{l+1,l}$, see~\cite{Lindblom98}, 
and the frequencies $\omega_i$ in the inertial and $\omega_c$ in the corotating frame are
\begin{align}\label{eq_rmd_freq_newt}
\omega_i &= - \frac{(l-1)(l+2)}{l+1} \Omega, &
\omega_c &= \frac{2\Omega}{l+1},
\end{align}
see~\cite{Pringle78}. 
The negative sign of $\omega_i$ means prograde motion of the wave patterns.

To measure the $r$-mode amplitude during a simulation, 
it is common to use a scalar product with the magnetic vector harmonic
times some radial weighting function.
This makes sense because the scalar product with other oscillation modes 
vanishes in the slow rotation limit.
A convenient choice to measure the $r$-mode amplitude is given
by the magnetic current multipole moment $J_{ll}$, 
which is also used to estimate the gravitational wave strain.
For the Newtonian $r$ mode, we find
\begin{align}
\delta J_{ll} 
  &= \int \Rmd r^l \delta\vec{v} \cdot  \vec{Y}^{B\star}_{ll}  \mathrm{d}^3x, \\
  &= \Half \Amp \Omega R^{1-l} e^{i\omega t} 
     \int \Rmd r^{2l} \left|\vec{Y}^B_{ll} \right|^2 \mathrm{d}^3x.  
\end{align}
Hence
\begin{align}
\Amp 
  &= \frac{\left|\delta J_{ll}\right|}{\Half \Omega R^{1-l}
        \int \Rmd r^{2l} \left|\vec{Y}^B_{ll} \right|^2 \mathrm{d}^3x }.  \label{eq_def_diml_ampl}
\end{align}
In this work, we use the above formula to define the dimensionless $r$-mode 
amplitude also for the rotating relativistic case, setting $R$ to the 
circumferential equatorial radius, and evaluating the denominator for the background model. 
Note this is not a covariant measure, 
for exact comparison to other works one should use
invariants like total energy or maximum velocity. 
With increasing rotation rate, one can expect that the presence of other modes 
(with the same $m$) starts contributing to the above measure as well.
This should not be a problem as long as the $r$ mode is the dominant one.
Differential rotation and other axisymmetric perturbations do not contribute to $J_{ll}$.

\subsection{Evolution equations}
We evolve the general relativistic hydrodynamic equations
for an ideal fluid, which in covariant form read
\begin{align}
  \nabla_\mu T^{\mu\nu} &= 0 \label{eq_div_tmunu},\\
  \nabla_\mu \left(\Rmd u^\mu \right) &=0.
\end{align}
The stress energy tensor $T^{\mu\nu}$ of an ideal fluid is given by
\begin{align}
  T^{\mu\nu} = \Rmd h u^\mu u^\nu + P g^{\mu\nu}.
\end{align}
For numerical evolution,
a 3+1-split is applied to obtain 
a first order system of hyperbolic evolution equations
in conservation form with source terms
\begin{align}
  \D_0 q &= -\D_i f^i(q,x^i) + s(q,x^i)  \label{eq_evol_cons}, \\
  q &\equiv \left(\Crmd,\Ced,\Cmom_j \right), 
\end{align}
with the evolved hydrodynamic variables given by
\begin{align}
  \Crmd   &\equiv \Vc\Lf \Rmd, \label{eq_def_crmd} \\
  \Ced    &\equiv \Vc \left( \Lf^2 \Rmd h - P -\Lf\Rmd \right),\\
  \Cmom_i &\equiv \Vc\Lf^2 \Rmd h v_i.\label{eq_def_cmom}
\end{align}
In flat Minkowski space (and using standard coordinates), 
$\Crmd$, $\Ced$, and $\Cmom^i$
reduce to mass density, energy density not including rest mass, 
and linear momentum density.
The flux terms $f^i=(f^i_{\Crmd},f^i_{\Ced},f^i_{\Cmom_j})$ are given by
\begin{align}
  f^i_{\Crmd}   &= w^i \Crmd,  \\
  f^i_{\Ced}    &= w^i \Ced  + \Lapse\Vc v^i P,  \\
  f^i_{\Cmom_j} &= w^i \Cmom_j  + \Lapse\Vc P \delta^i_j,
\end{align}
where $w^i = \Lapse v^i - \Shift^i$ is the advection speed relative to the coordinate system.
The source terms can be written in many ways, 
the formulation we are using is discussed in~\cite{Kastaun06}.
Finally, 
the evolution equations need to be completed by an equation of state (EOS) of the form
$P=P(\Rmd, \Sed)$ 
to compute the pressure.
\subsection{Conserved quantities}\label{sec_cons}
Making the assumption of a fixed axisymmetric spacetime 
not only simplifies the numerical evolution,
it also implies the existence of conserved fluxes.
The stationarity of the metric leads to a conserved energy density,
while the the axisymmetry of the metric
leads to a conserved angular momentum.
Conserved mass, energy, and angular momentum are given by
\begin{align}
M &= \int \Crmd \mathrm{d}^3x, &
E &= \int \Ped  \mathrm{d}^3x, &
J &= \int \Pam  \mathrm{d}^3x, \label{eq_def_cons}
\end{align}
where $\Crmd$ is defined by \Eref{eq_def_crmd} and,
using coordinates for which $\D_t, \D_\phi$ are Killing vectors,
\begin{align}
\Ped 
  &= \Vc \, T_{0\nu} n^\nu - \Crmd \\
  &= \Crmd \left( \Lf \Sent \Lapse -1 
     - \frac{\Lapse\Press}{\Lf\Rmd}  \right) - \Shift^i \Cmom_i, \label{eq_def_ped}\\  
\Pam 
  &= \Cmom_\phi 
  = -\Vc \, T_{\phi\nu} n^\nu 
  = \Crmd \Lf \Sent v_\varphi. 
\end{align}
Above, $n^\nu$ denotes the unit normal to the surfaces of constant coordinate time.
The conservation of $M,E,J$ can then be easily derived from the evolution
equations \eqref{eq_evol_cons}, 
without using the Einstein equations.
For an artificially fixed spacetime, 
conservation thus holds even for non-axisymmetric, non-stationary fluid flows.
The only requirement is that the fixed background metric is 
stationary and axisymmetric.

Note that $E$ depends on the gauge quantities $\Lapse$ and $\Shift$.
Since we require that $\D_t$ is a Killing vector, 
the only freedom of choice for $\Shift$ is given by translations along
space-like Killing vectors. 
The resulting change amounts to adding multiples of the conserved quantities
associated with those, e.g.\ angular momentum for a rotation.
The only freedom of choice for $\Lapse$ is multiplication by a constant,
which results in replacing $E$ by a linear combination with $M$. 
For the rest of this article, 
we assume asymptotic flatness 
and define $E$ as the energy computed for a coordinate frame
where $\Shift^i \to 0$ and $\Lapse \to 1$ at spatial infinity.
When making this choice, 
$\Ped \to 0$ for an infinitely diluted fluid element at rest at infinity, 
and hence a system becomes unbound for $E > 0$.

Also note that the numerically evolved energy density $\Ced$ is not the same
as the conserved energy density $\Ped$. 
In the Newtonian limit for example, 
for a fluid moving in a constant external gravitational field, 
$\Ced$ does not contain potential energy, in contrast to $\Ped$.
The reason for not using $\Ped$ in numerical schemes 
is that the corresponding evolution equation 
involves the time derivative of the gauge quantity $\Lapse$ 
in the fully relativistic case.

We stress that only $M$ is still conserved in full GR, 
and $E$ is not even conserved in Newtonian physics when allowing 
the gravitational field to change.
Further, $E+M$ is different from the ADM and Komar mass.
Interestingly,
$J$ equals the Komar angular momentum, 
as long as there are no singularities to consider.

\subsection{Oscillation energy}\label{sec_mode_energy}
In the following, we introduce a notation for the energy of oscillation modes
in the Cowling approximation, 
which is useful as a diagnostic tool in numerical studies performed in a fixed spacetime,
but might also serve as an order of magnitude estimate in the general case.

For this we expand the conserved energy $E$ defined in the previous section 
around a stationary background model.
For non-axisymmetric modes, it is easy to show that the linear terms cancel 
when integrating along the $\phi$-direction,
so we need to expand the energy to second order in the amplitude.
Since we want an expression that can be computed without knowledge
of the second order corrections to the eigenfunction,
we still assume that the perturbation itself scales linearly.
However, this leads to conceptional problems.
First, there is a freedom of choice concerning the set of variables used
to completely specify the system.
If those variables are perturbed linearly,
the resulting energy perturbation at second order depends on this choice.
Second, we found that in general the conserved mass changes as well 
when evaluated to second order. 
This implies that the state around which the model oscillates is not exactly the
same as the unperturbed state.
Unfortunately, we are interested in the energy difference from 
the stationary state reached after all the oscillation energy is dissipated,
e.g.\ due to numerical damping, 
while conserving total mass and angular momentum profile.
For one of our models, we computed mass and energy change when perturbing
$\Rmd, v^i$ with the $r$-mode eigenfunction.
By assuming that the mass is created with the average binding energy $E/M$ of 
the background model,
we estimated the ambiguity in defining $E$ to be around 50 \%.
This effect could be tracked down to the fact that not only the kinetic energy density,
but also the conserved mass density depend on the velocity via the Lorentz factor, 
in contrast to the Newtonian case.

To cure these problems, we define the oscillation energy as the perturbation
of the conserved energy when perturbing the variables $\Crmd, \Pam, v^a$ linearly.
This way, total mass and angular momentum are exactly conserved for non-axisymmetric
perturbations.
To our knowledge, the following has not been discussed elsewhere.
We will assume a cylindrical coordinate system adapted to the symmetries 
and with $\Shift^a = g_{a\phi} = 0$,
and only consider background models with $v^a=0$.
For a non-axisymmetric mode described by \Eref{eq_generic_mode}, 
we define the mode energy as
\newcommand{\upr}{\bar{U}}
\newcommand{\fpr}{\bar{f}}
\newcommand{\pd}[2]{\frac{\D #1}{\D#2}}
\newcommand{\spd}[2]{\frac{\D^2 #1}{\D#2^2}}
\newcommand{\mpd}[3]{\frac{\D^2 #1}{\D #2 \D #3}}
\begin{align}
\begin{split}
\hat{E} 
  &= \frac{1}{\Amp^2}\int \left(  
      \Half \spd{\Ped}{\Crmd} \delta\Crmd^2
      +\Half \spd{\Ped}{\Pam} \delta\Pam^2                   \right. \\ & \quad \left. 
      + \mpd{\Ped}{\Crmd}{\Pam} \delta\Crmd \delta\Pam
      + \Half \mpd{\Ped}{v^a}{v^b} \delta v^a \delta v^b 
     \right) \mathrm{d}^3x         \label{eq_def_mode_energy}
\end{split}\\
\begin{split}
  &= \pi \int \left(
      \Half \spd{\Ped}{\Crmd} \hat{\Crmd}^2
      +\Half \spd{\Ped}{\Pam} \hat{\Pam}^2                   \right. \\ & \quad \left. 
      + \mpd{\Ped}{\Crmd}{\Pam} \hat{\Crmd} \hat{\Pam}
      + \Half \mpd{\Ped}{v^a}{v^b} \hat{v}^a \hat{v}^b 
\right) \mathrm{d}\vartheta \, \mathrm{d}z. \label{eq_def_mode_energy_cyl}
\end{split}
\end{align}
The quantities in \Eref{eq_def_mode_energy}  are defined with respect to Cartesian coordinates, 
while \Eref{eq_def_mode_energy_cyl} is valid in cylindrical coordinates.
Terms with $\delta\Crmd \delta v^a$ or $\delta\Pam \delta v^a$
do not contribute since they have the angular dependency  
$\sin(m\phi + \omega t) \cos(m\phi + \omega t)$.
Also the corresponding second derivatives are zero.
This energy depends on the normalization of the eigenfunctions, 
which has to be specified.
The eigenfunctions of the conserved variables are given by
\begin{align}
\hat{\Crmd}
  &= \Vc \Lf \left( \hat{\Rmd} + \Rmd \Lf^2 v_\phi \hat{v}^\phi \right), \\
\hat{\Pam}
  &= \Vc \Lf^2 \Sent \left[ \left(1 + \Csnd^2 \right) v_\phi \hat{\Rmd} 
       + \Rmd \left(2 \Lf^2 - 1 \right) \Gpp \hat{v}^\phi \right].
\end{align}
To compute the second derivatives, we first define
\begin{align}
\upr(\Lf, \Rmd, \Pam) 
  &= -\Shift^\phi \Pam + \Vc\left( 
     \Lf \Rmd \left(\Lf \Sent \Lapse -1 \right) - \Lapse \Press \right) \label{eq_def_upr}\\
  &= \Ped(\Crmd,\Pam, v^a).
\end{align}
We then write
\begin{align}
\begin{split}
\spd{\Ped}{\Crmd} 
  &= \spd{\upr}{\Lf} \left( \pd{\Lf}{\Crmd} \right)^2
     + \pd{\upr}{\Lf} \spd{\Lf}{\Crmd}
     + \pd{\upr}{\Rmd} \spd{\Rmd}{\Crmd} \\ &\quad 
     + \spd{\upr}{\Rmd} \left( \pd{\Rmd}{\Crmd} \right)^2
     + 2 \mpd{\upr}{\Lf}{\Rmd} \pd{\Rmd}{\Crmd} \pd{\Lf}{\Crmd},
\end{split}
\\
\begin{split}
\spd{\Ped}{\Pam} 
  &= \spd{\upr}{\Lf} \left( \pd{\Lf}{\Pam} \right)^2
     + 2 \mpd{\upr}{\Lf}{\Rmd} \pd{\Rmd}{\Pam} \pd{\Lf}{\Pam} \\ &\quad
     + \spd{\upr}{\Rmd} \left( \pd{\Rmd}{\Pam} \right)^2
     + \pd{\upr}{\Lf} \spd{\Lf}{\Pam}
     + \pd{\upr}{\Rmd} \spd{\Rmd}{\Pam} ,
\end{split}
\\
\begin{split}
\mpd{\Ped}{\Crmd}{\Pam} 
  &= \spd{\upr}{\Lf} \pd{\Lf}{\Crmd} \pd{\Lf}{\Pam}
     + \spd{\upr}{\Rmd} \pd{\Rmd}{\Pam} \pd{\Rmd}{\Crmd} \\ &\quad
     + \mpd{\upr}{\Lf}{\Rmd} \left( 
        \pd{\Rmd}{\Pam} \pd{\Lf}{\Crmd} + \pd{\Lf}{\Pam} \pd{\Rmd}{\Crmd} 
       \right)  \\ &\quad
     + \pd{\upr}{\Lf} \mpd{\Lf}{\Crmd}{\Pam}
     + \pd{\upr}{\Rmd} \mpd{\Rmd}{\Crmd}{\Pam} ,
\end{split}
\\
\mpd{\Ped}{v^a}{v^b}
  &= \pd{\upr}{\Lf} \mpd{\Lf}{v^a}{v^b} + \pd{\upr}{\Rmd} \mpd{\Rmd}{v^a}{v^b},
\end{align}
where we included only nonzero terms.
From \Eref{eq_def_upr}, we compute
\begin{align}
\pd{\upr}{\Lf}
  &= \Vc \Rmd \left( 2\Lapse \Sent \Lf  - 1 \right), \\
\pd{\upr}{\Rmd}
  &= \Vc \Lf \left(\Lapse \Sent \Lf \left( 1 + v^2 \Csnd^2 \right)  - 1 \right), \\
\spd{\upr}{\Lf}
  &= 2 \Vc \Lapse \Rmd \Sent, \\
\spd{\upr}{\Rmd}
  &= \Vc \Lapse \Lf^2 \Sent \left( \frac{\Csnd^2}{\Rmd}\left( 1 + v^2\Csnd^2 \right) 
        + v^2 \pd{}{\Rmd} \Csnd^2 \right), \\
\mpd{\upr}{\Lf}{\Rmd}
  &= \Vc \left( 2 \Lapse \Sent \Lf \left( 1 + \Csnd^2 \right) - 1 \right).
\end{align}
The functions $\Lf(\Crmd,\Pam, v^a), \Rmd(\Crmd,\Pam, v^a)$ 
cannot be expressed in closed analytic form.
However, we only need the derivatives. 
By computing the derivatives of the conserved variables with respect
to the primitives and then inverting the resulting linear system of equations,
we arrive at
\begin{align}
\pd{v^\phi}{\Crmd}
  &= - \frac{1+\Csnd^2}{\Vc \Lf \Rmd f} v_\phi \label{eq_dvp_dcrmd},\\
\pd{v^\phi}{\Pam}
  &= \frac{1}{\Vc \Lf^2 \Rmd \Sent f}, \\
\pd{\Rmd}{\Crmd}
  &= \frac{2\Lf^2 -1}{\Vc\Lf f} \Gpp, \\
\pd{\Rmd}{\Pam}
  &= - \frac{v_\phi}{\Vc \Sent f},  \label{eq_drmd_dpam}
\end{align}
where
\begin{align}
f &= \Lf^2 v_\phi^2 \left( 1 - \Csnd^2 \right) + \Gpp. 
\end{align}
It follows that
\begin{align}
\pd{\Lf}{\Crmd}
  &= - \frac{\Lf^2 v^2}{\Vc \Rmd f} \left( 1 + \Csnd^2 \right) \Gpp, \\
\pd{\Lf}{\Pam}
  &= \frac{\Lf}{\Vc \Rmd \Sent f} v_\phi. \label{eq_dlf_dpam}
\end{align}
Using \Erefs{eq_dvp_dcrmd}{eq_dlf_dpam}, we obtain
\begin{align}
\begin{split}
\pd{f}{\Crmd}
  &= - \frac{\Lf v^2 \Gpp^2}{\Vc f} \left[
            2 \frac{\Lf^2}{\Rmd} \left( 1 - \Csnd^4 \right)             \right.  \\ &\quad  \left.
            + \left( 2 \Lf^2 -1 \right) \pd{}{\Rmd} \Csnd^2 \right] ,
\end{split} \\
\pd{f}{\Pam}
  &= v_\phi \frac{\Lf^2 \Gpp}{\Vc \Sent f}  \left[ \frac{2}{\Rmd} \left( 1 - \Csnd^2 \right) 
             + v^2 \pd{}{\Rmd} \Csnd^2 \right].
\end{align}
Finally, we find
\begin{align}
\begin{split}
\spd{\Lf}{\Crmd}
  &=  \left[ 
      \left( 1 + \Csnd^2 \right) 
      \left( \frac{v^2}{f}\pd{f}{\Crmd} - 2 \Lf^2 v_\phi \pd{v^\phi}{\Crmd} \right)     \right.\\&\quad\left.
      +v^2 \left(\frac{1+\Csnd^2}{\Rmd} - \pd{}{\Rmd} \Csnd^2 \right) \pd{\Rmd}{\Crmd} 
      \right] \frac{\Lf^2 \Gpp}{\Vc \Rmd f},
\end{split} \\
\begin{split}
\spd{\Lf}{\Pam}
  &= \frac{\Lf}{\Vc \Rmd \Sent f} \left[
      \Lf^2 \Gpp \pd{v^\phi}{\Pam}                      \right.  \\ &\quad  \left.
      - \frac{v_\phi}{f} \pd{f}{\Pam} 
      - \frac{v_\phi}{\Rmd} \left(1 + \Csnd^2 \right) \pd{\Rmd}{\Pam} \right],
\end{split} \\
\begin{split}
\mpd{\Lf}{\Pam}{\Crmd}
  &= \frac{\Lf}{\Vc \Rmd \Sent f} \left[
      \Lf^2 \Gpp \pd{v^\phi}{\Crmd}                      \right.  \\ &\quad  \left.
      - \frac{v_\phi}{f} \pd{f}{\Crmd} 
      - \frac{v_\phi}{\Rmd} \left(1 + \Csnd^2 \right) \pd{\Rmd}{\Crmd} \right],
\end{split}\\
\mpd{\Lf}{v^a}{v^b}
  &=  \frac{\Lf^3}{f}\Gpp g_{ab},
\end{align}
\begin{align}
\begin{split}
\spd{\Rmd}{\Crmd}
  &= \frac{\Gpp}{\Vc f} \left[
       \Lf \left(2 \Lf^2 + 1 \right) v_\phi \pd{v^\phi}{\Crmd}             \right.  \\ &\quad  \left.
       - \left( 2\Lf^2 - 1 \right) \frac{1}{\Lf f} \pd{f}{\Crmd} \right], 
\end{split}\\
\spd{\Rmd}{\Pam}
  &= \frac{1}{\Vc h f} \left[
       \frac{v_\phi}{f} \pd{f}{\Pam} - \Gpp \pd{v^\phi}{\Pam} 
       + v_\phi \frac{\Csnd^2}{\Rmd} \pd{\Rmd}{\Pam} \right], \\
\mpd{\Rmd}{\Pam}{\Crmd}
  &= \frac{1}{\Vc h f} \left[
       \frac{v_\phi}{f} \pd{f}{\Crmd} - \Gpp \pd{v^\phi}{\Crmd} 
       + v_\phi \frac{\Csnd^2}{\Rmd} \pd{\Rmd}{\Crmd} \right], \\
\mpd{\Rmd}{v^a}{v^b}
  &= - \frac{\Gpp}{f} \Rmd \Lf^2  g_{ab}.        
\end{align}
First and second order derivatives have of course been computed without assuming
$v^a=0$, but the results given here have been evaluated at $v^a=0$ for
simplicity.

We stress that the energy defined by \Eref{eq_def_mode_energy} is only an estimate for 
the energy of a finite amplitude oscillation.
It was shown in~\cite{Sa2004} that for $r$ modes of Newtonian stars, 
differential rotation is an unavoidable feature.
Most likely, such axisymmetric terms would contribute 
to the first order expansion of the energy $E$
and hence constitute a second-order contribution in total, 
like the terms considered in our definition.
Without knowledge of the second-order perturbation, 
\Eref{eq_def_mode_energy} is probably the best one can do.

%
\section{Numerical method}\label{sec_meth}
In the following we briefly describe our numerical methods. 
For readers not familiar with general relativistic hydrodynamics,
we recommend the review~\cite{FontLRR.V7}.

\subsection{Time evolution}

We evolve the 3+1 split hydrodynamic evolution equations in 
flux-conservative form (\ref{eq_evol_cons}).
However, we use a zero-temperature EOS of the form $P=P(\Rmd)$.
As a consequence, the system becomes overdetermined.
Therefore, we do not evolve the energy density $\Ced$, which becomes redundant,
but compute it from mass and momentum densities. 
For details, see~\cite{Kastaun06}.
We stress that this approach is only self-consistent for adiabatic evolution,
and hence our simulations are only correct in the absence of shock formation.
Discontinuities cannot produce shock heating, 
instead they lead to a violation of energy conservation.
A sharp decrease of $E$ is therefore an indicator for shock formation,
see~\cite{Kastaun10} for examples.

In the absence of shocks however,
the evolution becomes more accurate,
since there is no error in the evolution of the specific entropy.
One particular error that is avoided this way is the formation of large scale, 
low velocity convective movements driven by entropy gradients caused by numerical errors.
Such vortices could easily be confused with genuine nonlinear effects
in simulations of high amplitude r mode oscillations.

To evolve the above system numerically, we use the \Pizza code first described 
in~\cite{Kastaun06}.
It is based on an HRSC (high resolution shock capturing) scheme,
which was optimized for quasi-stationary simulations.
As shown in~\cite{Kastaun06}, the code is able to evolve a stationary star 
with high accuracy, in particular the rotation profile.
As for all such codes, special care has to be taken to treat the stellar surface.
Instead of using an artificial atmosphere, we apply the scheme used
in~\cite{Kastaun10}. The advantages are that the mass is conserved to
machine precision and that the amplitude of oscillations excited by numerical noise
is negligible for the applications presented here.
Still, the surface treatment is the main source of numerical damping.

We use uniform three-dimensional Cartesian grids 
in co-rotating coordinates.
The outer boundaries are placed far enough from the surface
to prevent matter from leaving the computational domain.
Thus, the total mass $M$ is conserved to machine precision.
The total angular momentum on the other hand is subject to discretization errors,
since $\Pam$ is not an evolved variable in Cartesian coordinates.

\subsection{Eigenfunction extraction}\label{sec_methods_ef}

In order to extract eigenfunctions numerically, 
we use the mode recycling method in the form described
in~\cite{Kastaun10}. 
In short, the star is perturbed using some trial perturbation,
and then evolved in time numerically.
The frequencies of the oscillations excited by the perturbation 
are determined using Fourier analysis in time at some sample point.
By selecting one frequency $\omega_0$ and computing the Fourier integrals
at this frequency for any point in the star, one obtains a first
estimate of the complex eigenfunction,
which is usually still contaminated by other oscillations due to the finite
evolution time.
To obtain the two-dimensional eigenfunctions, we divide
the numerical complex eigenfunction by $e^{im\phi}$ and average over $\phi$.
This removes contributions with different $\phi$-dependency.
Next, we remove contributions with the wrong $z$-parity by (anti-)symmetrizing. 
Finally, we compute and factor out the average complex phase (see~\cite{Kastaun10} for details),
while using the phase variance as an error measure.
The result is used as the initial perturbation in a new simulation,
and the whole process is repeated until oscillation modes other than the desired one
are reasonably suppressed.

Obviously, the method is only effective if the initial trial perturbation
significantly excites the desired mode.
To extract the $r$-mode eigenfunction,  
we choose the Newtonian eigenfunction for a slowly rotating star,
given by \Eref{eq_ef_r_newt}, 
which is close enough to the actual eigenfunctions of the models. 

\section{Stellar models}
\label{sec_models}
We investigate two different uniformly rotating stellar models with fixed central density 
$\rho_c = 7.9053\e{17} \usk\kilogram\,\meter\rpcubed$
and EOS,
but different rotation rates.
Their properties are summarized in \Tab{tab_models}. 
The EOS, which is also used during the evolution,
is a polytropic EOS defined by
\begin{align}
  P(\Rmd) = \Rmdp\left(\frac{\Rmd}{\Rmdp}\right)^\Gamma,
  \label{eq_eos_poly}
\end{align}
with polytropic exponent $\Gamma=2$ and 
the constant density scale $\Rmdp = \unit{6.1760\e{18}}{\kilogram\usk\meter\rpcubed}$.
We note that polytropic stars are stable against convection
and do not possess $g$ modes, 
i.e. modes for which buoyancy is the restoring force.
Model \BUthree\ was computed using the \textsc{rns} code described in~\cite{RNScode},
while model \BUsix\ was computed using the code described in~\cite{Ansorg03,Ansorg08}.
Both codes are accurate enough for our purposes, 
the only reason to use different codes is that the latter became our standard choice
once it was available.

These models are a crude approximation to real neutron stars.
Their purpose is to get a basic qualitative understanding of 
the nonlinear r mode dynamics in the most simple case.
It is likely that the inclusion of composition gradients
or differential rotation will lead to new effects.

To excite nonlinear oscillations, 
we always use the exact eigenfunction obtained by mode recycling.
However, in contrast to the linear regime, 
there is some arbitrariness involved regarding how 
to scale the eigenfunctions to large amplitudes.
Ideally, we would like initial data resembling 
a mode naturally grown to high amplitudes, 
e.g.\ due to the CFS mechanism.
Since this is not feasible, we simply scale the perturbation of the velocity and of the 
specific energy and recompute the other fluid quantities consistently after applying the perturbation.

Note also that we use Eulerian perturbations. 
This has the drawback that the star is not deformed correctly close to the surface,
in particular there is no perturbation outside the surface of the unperturbed star.
For high amplitudes, the initial data inevitably contains small shocks at the surface.
In practice however, even high amplitude r modes do not induce large deformations 
and we did not notice the corresponding numerical artifacts.

\begin{table}
\begin{tabular}{l|l|l|l|l|l|l}
Name & $M_B / M_\odot$ & $F_R / \hertz$ & $R_c / \kilo\meter$ & $a_r$ & $e_c$ & $\beta_c$ \\\hline
\BUthree  & 1.6194          & 590.90         & 15.384     & 0.85  &  -0.2094  & 0.02406 \\
\BUsix    & 1.7555          & 792.10         & 17.268     & 0.7   &  -0.2159  & 0.04864 \\
\end{tabular}
\caption{\label{tab_models}
Details of the stellar models. 
$M_B$ is the total rest mass, 
$F_R$ the rotation rate as observed from infinity,
$R_c$ the equatorial circumferential radius, 
$a_r$ the ratio of polar to equatorial coordinate radius, 
$e_c = E / M_B$,
$\beta_c = |\Omega J / (2E)|$, where $E$ and $J$ are energy and angular momentum 
defined by \Eref{eq_def_cons}.
}
\end{table}

\section{Numerical results}
\label{sec_res}
In the following, we present our results for the $r$ mode with $l=m=2$.
Unless noted otherwise, 
our simulations use a uniform resolution of 50 points per equatorial stellar radius,
which is a reasonable compromise between accuracy and
computational cost.

\subsection{R-mode properties}\label{sec_mode_prop}
Using the methods in \Sec{sec_methods_ef},
we extracted eigenfunctions and frequencies of the  $r$ mode
for the models in \Tab{tab_models}.
We also computed the energy of the modes defined in \Sec{sec_mode_energy}.
The results are given in \Tab{tab_freq_rmd}.
For our models, the frequencies agree with the ones in the Newtonian 
slowly rotating case better than 10 \%.
The $r$-mode frequencies in the inertial frame 
found by \cite{Gaertig2010} agree with our results
better than 0.1\%. 
Given that we use a completely independent code
based on a different method, 
the good agreement validates the results.

The mode energy is useful to quantify what amplitudes are large
in the sense that strong deformations of the star occur.
Naively, one should think that e.g.\ $\Amp=3$ is a huge amplitude
because the velocity perturbations become comparable to the
rotational velocity at the equator.
Looking at the energy however, 
we find that the mode energy at $\Amp=3$ 
for model \BUthree\ is only a fraction $3\e{-3}$ of the binding energy $E$.
For model \BUthree, 
the energy of the perturbation equals the stars' binding energy
at $\Amp \approx 57$.
For the mode recycling process, we used amplitudes around $\Amp=0.3$,
well inside the linear regime.

\begin{table}
\begin{tabular}{l|l|l|l|l}
Model     & $f_c / \hertz$ & $f_i / \hertz$ & $f_c / f_c^N$ & $\hat{E} / E$  \\\hline
\BUthree  & 361.4          & 820.4          & 0.917         & 3.033\e{-4}     \\
\BUsix    & 544.8          & 1039.4         & 1.032         & 5.093\e{-4}    \\
\end{tabular}
\caption{\label{tab_freq_rmd}
Properties of the $l=m=2$ $r$-mode. 
$f_c$ is the frequency with respect to the co-rotating frame, 
$f_i$ the frequency observed from infinity in the inertial frame. 
The estimated numerical accuracy of $f_c$ is 1\%, 
not including the unknown error due to the Cowling approximation.
For comparison we compute the $r$-mode frequency $f_c^N$ for slowly 
rotating Newtonian stars given by \Eref{eq_rmd_freq_newt}.
$\hat{E}$ is the $r$-mode energy, defined by \Eref{eq_def_mode_energy},
normalized to an amplitude $\Amp=1$ as defined by \Eref{eq_def_diml_ampl}.
$E$ is the models' binding energy defined by \Eref{eq_def_cons}.
}
\end{table}

\begin{figure}
\includegraphics[width=\columnwidth]{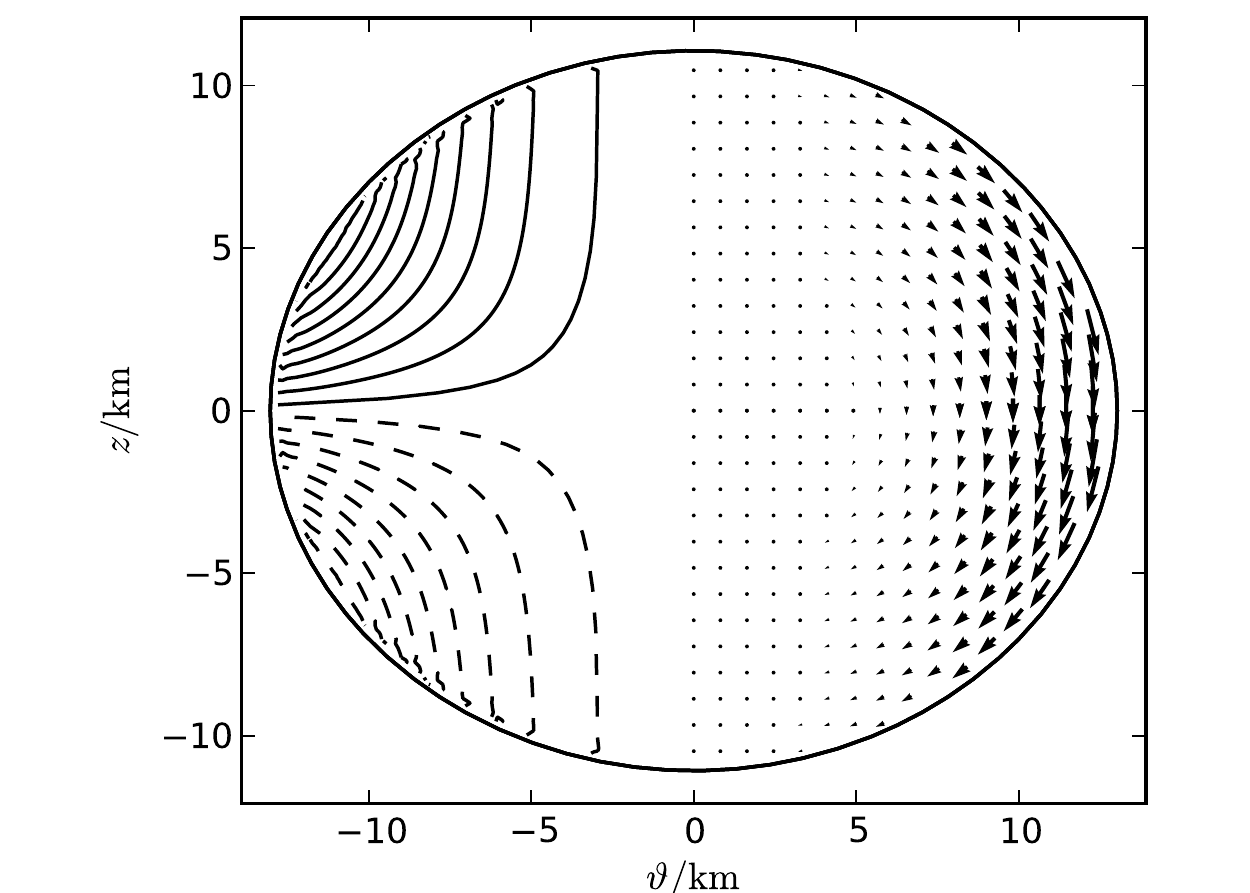}
\caption{\label{fig_bu3_ef}
Two-dimensional eigenfunctions $\hat{\epsilon}(\vartheta,z)$ (left half) 
and $\hat{v}^a(\vartheta,z)$ (right half), 
belonging to the $r$ mode of model \BUthree. 
}
\end{figure}

\begin{figure}
\includegraphics[width=\columnwidth]{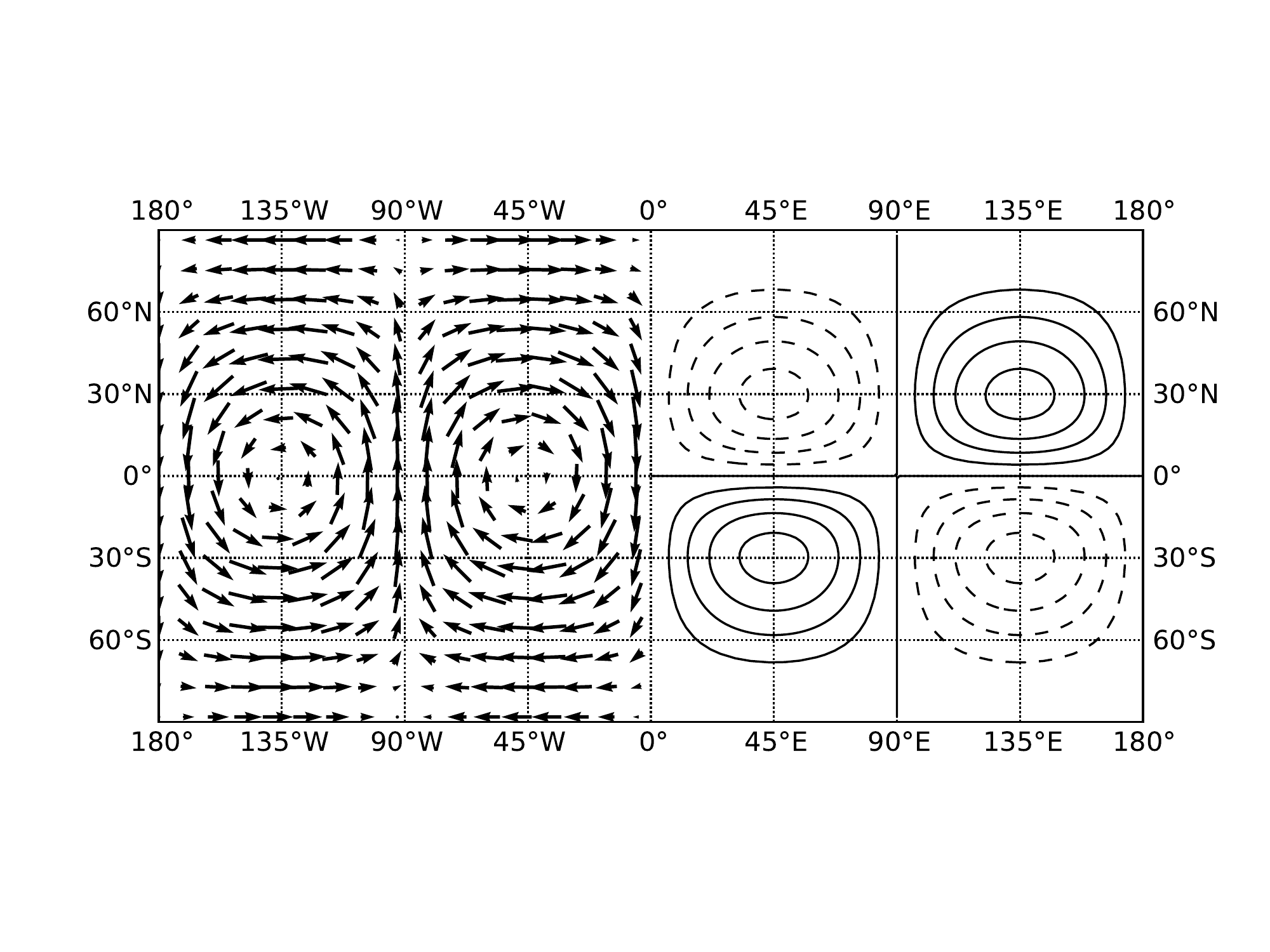}
\caption{\label{fig_bu3_ef_sph}
Perturbations corresponding to the $r$ mode of model \BUthree,
as longitude/latitude plot at fixed coordinate radius $r=R_e/2$,
with $R_e$ being the equatorial coordinate radius.
The velocity perturbations $\delta v^\phi, \delta v^\theta$
are plotted on the left half, and the specific energy perturbation
$\delta \Sed$ on the right half.
Note the patterns have a 180\degree-periodicity in $\phi$. 
}
\end{figure}

\begin{figure}
\includegraphics[width=\columnwidth]{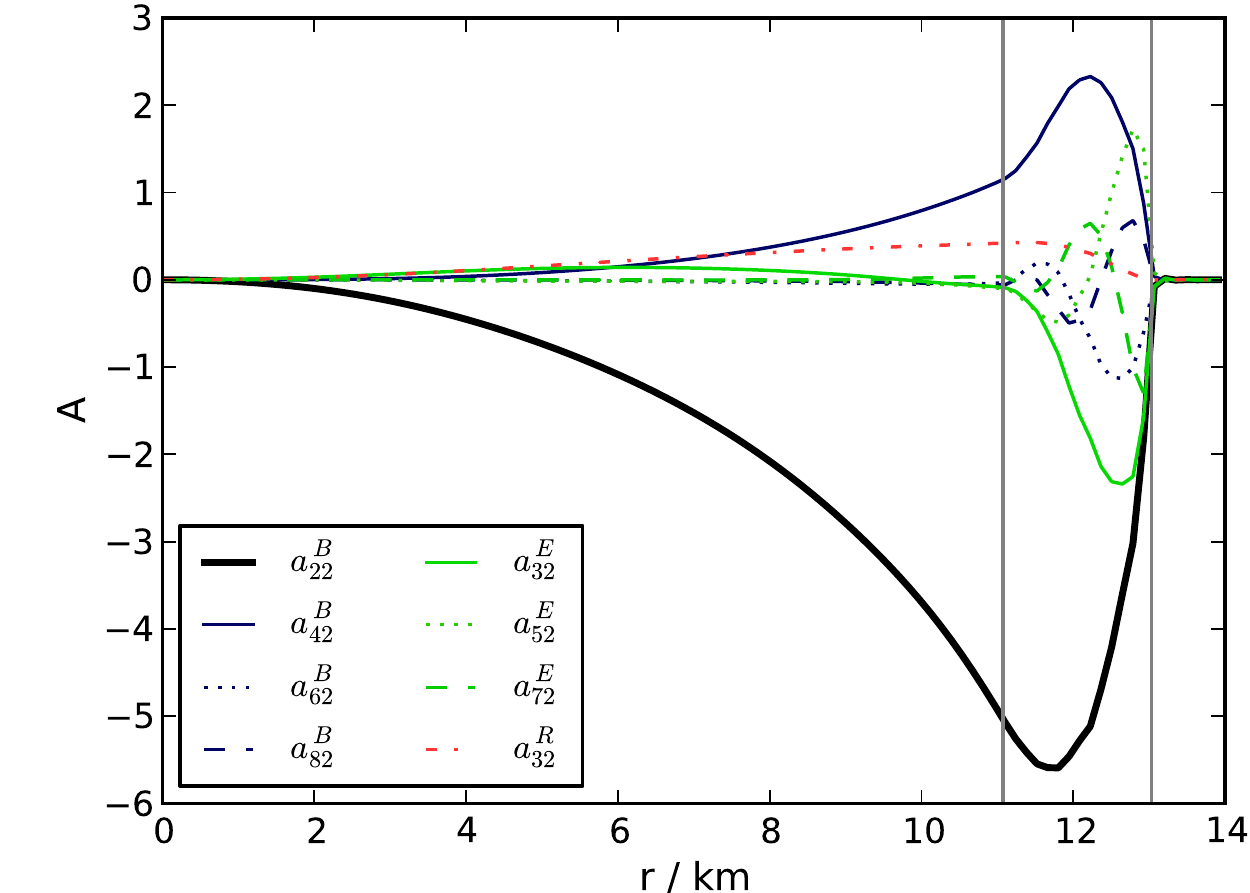}
\caption{\label{fig_bu3_ef_mp}
Decomposition of the $r$-mode velocity perturbation of model 
\BUthree\ into vector spherical harmonics. 
The vertical lines mark the polar and equatorial radius.
}
\end{figure}

The eigenfunctions are visualized in Figs.~\ref{fig_bu3_ef}--\ref{fig_bu3_ef_sph}.
As one can see, 
the velocity perturbations are similar to those in the Newtonian slowly rotating
case, in particular the radial component is small. 
Also the density perturbation is qualitatively the same as in the Newtonian case.
To quantify the differences, we decompose the velocity perturbation into vector 
spherical harmonic functions
\begin{align}
\begin{split}
\delta \vec{v} 
  &= \sum_{l=0}^\infty \sum_{m=-l}^l \left( 
      a_{lm}^E(r) \vec{Y}^E_{lm}(\theta,\phi)   \right. \\ &\quad \left.
      + a_{lm}^B(r) \vec{Y}^B_{lm}(\theta,\phi)
      + a_{lm}^R(r) Y_{lm}(\theta,\phi)\hat{e}_r
     \right).
\end{split}
\end{align}
The results are plotted in \Fig{fig_bu3_ef_mp}. 
The dominant contribution to the velocity is the
magnetic type vector harmonic. 
However, there are also significant radial and polar contributions.
The dominant component of the specific energy (not shown in the plot) 
is the $l=3, m=2$ spherical harmonic.

Although higher order multipole moments are quite small in the inner regions
of the star, they have significant amplitudes between the polar and equatorial radius.
This does not imply that small scale structures appear close to the surface.
The reason is just that the star is ellipsoidal and the spheres of constant
coordinate radius start intersecting with the stellar surface. 

From the multipole decomposition of the numerically extracted eigenfunctions,
we estimated the gravitational radiation caused by the $r$ modes,
using the multipole formulas for Newtonian sources from~\cite{Thorne80}.
Luminosity, strain, and angular momentum loss are given in \Tab{tab_gw_rad}.
Unsurprisingly, we find that for the $r$ modes only the $l=m=2$ current multipole 
contributes significantly. 
The second largest contribution comes from the $l=3,m=2$ mass multipole,
which for model \BUthree\ is smaller by a factor 
$|A^{E2}_{32} / A^{B2}_{22}| \approx 0.02$. 
The higher multipole moments are completely negligible.

\begin{table}
\begin{tabular}{l|l|l|l}
Model     & $A^{B2}_{22} / (10\usk\mega\parsec) $  & $\tau_E / \second$ & $\tau_J / \second $  \\\hline
\BUthree  & $1.42\e{-24}$         &$1.47\e{6}$      & $4.93\e{4}$       \\
\BUsix    & $3.56\e{-24}$         &$1.64\e{5}$      & $1.04\e{4}$       \\
\end{tabular}
\caption{\label{tab_gw_rad}
Gravitational radiation caused by $r$ modes in the linear regime.
The gravitational luminosity $W_{gw}$ and angular momentum loss $\dot{J}_{gw}$ 
are given in terms of the timescales
$\tau_J = J / \dot{J}_{gw}$ and $\tau_E = |E| / W_{gw}$,
where $E$ and $J$ are the binding energy and angular momentum given in \Tab{tab_models}.
The values are normalized to an amplitude $\Amp=1$, 
with  $A^{B2}_{22} \sim \Amp$, $W_{gw} \sim \Amp^2$, $\dot{J}_{gw} \sim \Amp^2$.
}
\end{table}

\subsection{Rotation profile}\label{sec_diffrot}
One of the most noticeable features in our simulations is the 
development of strong differential rotation during the evolution of 
$r$ modes with high initial amplitudes.

To visualize the rotation profile alone without the contribution from the 
$r$-mode oscillation, we compute the $\phi$-averaged angular velocity.
Of course, any axisymmetric oscillation would contribute to this measure as well.
However, since the Fourier spectrum of the $\phi$-velocity at some sample points
in the star did only show significant peaks at the $r$-mode frequency and at zero
frequency,
we can assume that any snapshot of the $\phi$-averaged angular velocity
is a good measure of the differential rotation.

\begin{figure}
\includegraphics[width=\columnwidth]{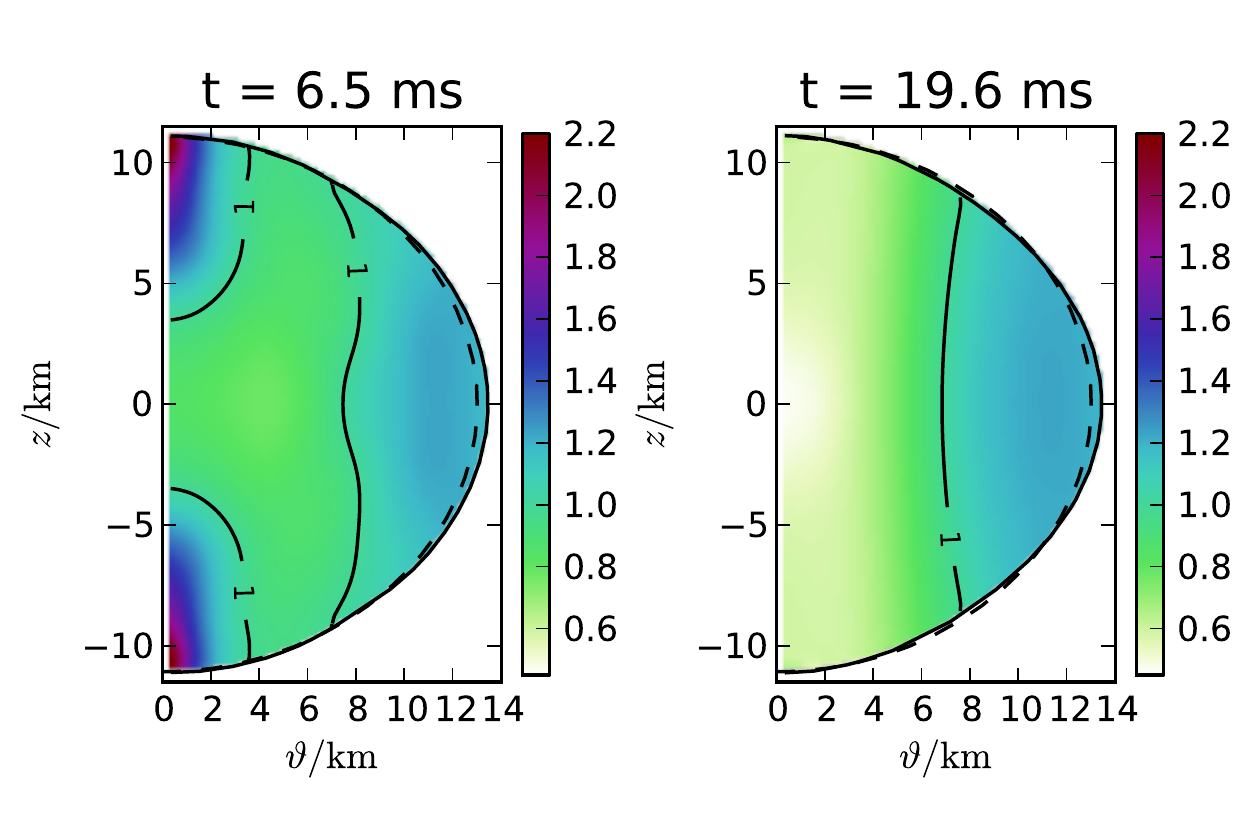}
\caption{\label{fig_bu3_diffrot}
Development of differential rotation, 
for model \BUthree\ perturbed with 
an $r$ mode of initial amplitude $\Amp=3$.
Color coded is the $\phi$-averaged angular velocity in units of 
the angular velocity of the unperturbed model.
The surface of the unperturbed star is marked by a thin dotted line,
while the surface at which the $\phi$-averaged density 
falls below 0.005 times the central density is marked by a thin solid line. 
A small region near the axis has been excluded to avoid problems when 
transforming from the Cartesian grid.
}
\end{figure}

\begin{figure}
\includegraphics[width=\columnwidth]{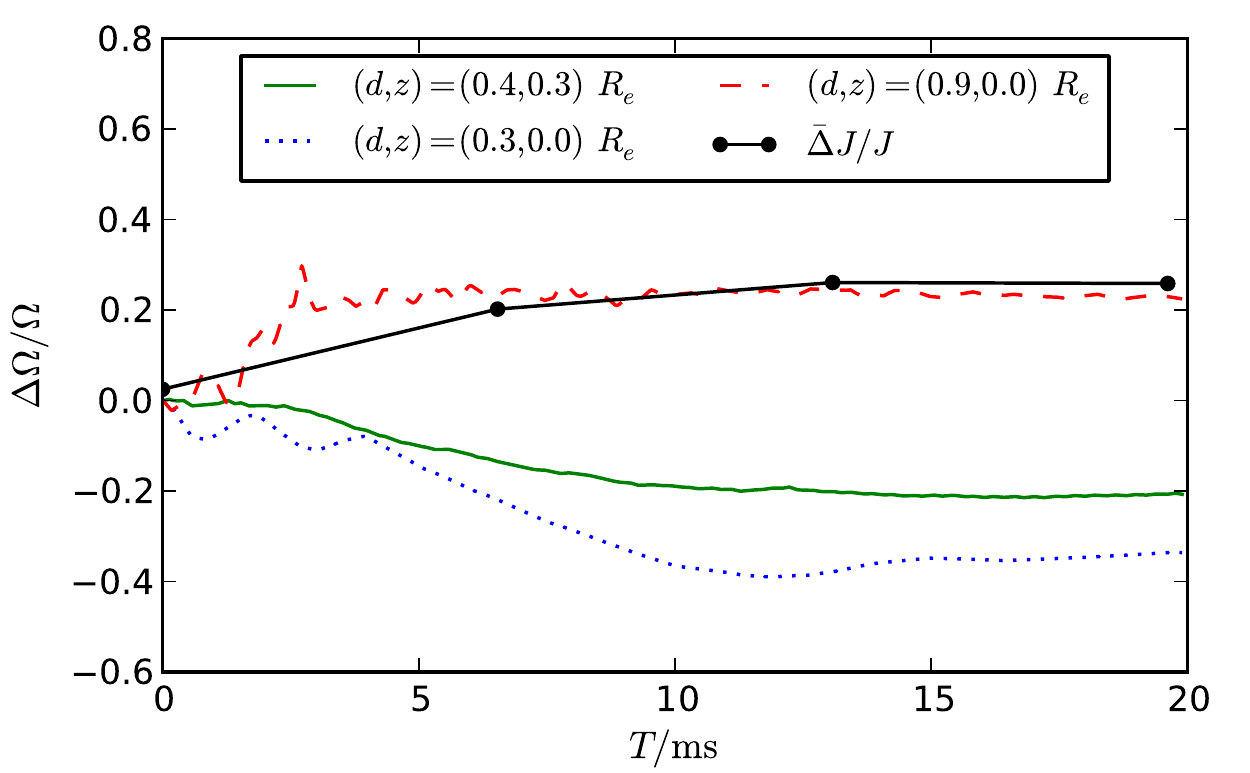}
\caption{\label{fig_bu3_diffrot_evol}
Time evolution for the same setup as \Fig{fig_bu3_diffrot} 
of the $\phi$-averaged angular velocity 
at different sample points in the star.
Also shown is the global average angular momentum change defined by 
\Eref{eq_avg_diffrot}. 
}
\end{figure}

\Fig{fig_bu3_diffrot} shows snapshots at two different times during and after 
the decay. 
During the decay, the angular velocity shows a two-dimensional structure.
Strangely, the angular velocity near the poles temporarily increases to more than twice
the initial value and then decreases again.
At a later stage, the angular velocity converges to a simple profile 
depending roughly on the distance 
to the axis. 
In order to quantify the amount of angular momentum redistribution,
we define an average change of angular momentum by
\begin{align}\label{eq_avg_diffrot}
\frac{\bar{\Delta} J}{J} 
  &= \sqrt{
      \frac{\int \left(\bar{L} - \bar{L}_0 \right)^2  \mathrm{d}^2x}
           {  \int \bar{L}_0^2  \mathrm{d}^2x}
     }, \\
  \bar{L} &= \frac{1}{2\pi} \int_0^{2\pi} L \, \mathrm{d}\phi,
\end{align}
where $L_0$ is the value for the unperturbed model.
Figure~\ref{fig_bu3_diffrot_evol} shows the time evolution of this measure 
as well as the local differential rotation at chosen positions.

\begin{figure}
\includegraphics[width=\columnwidth]{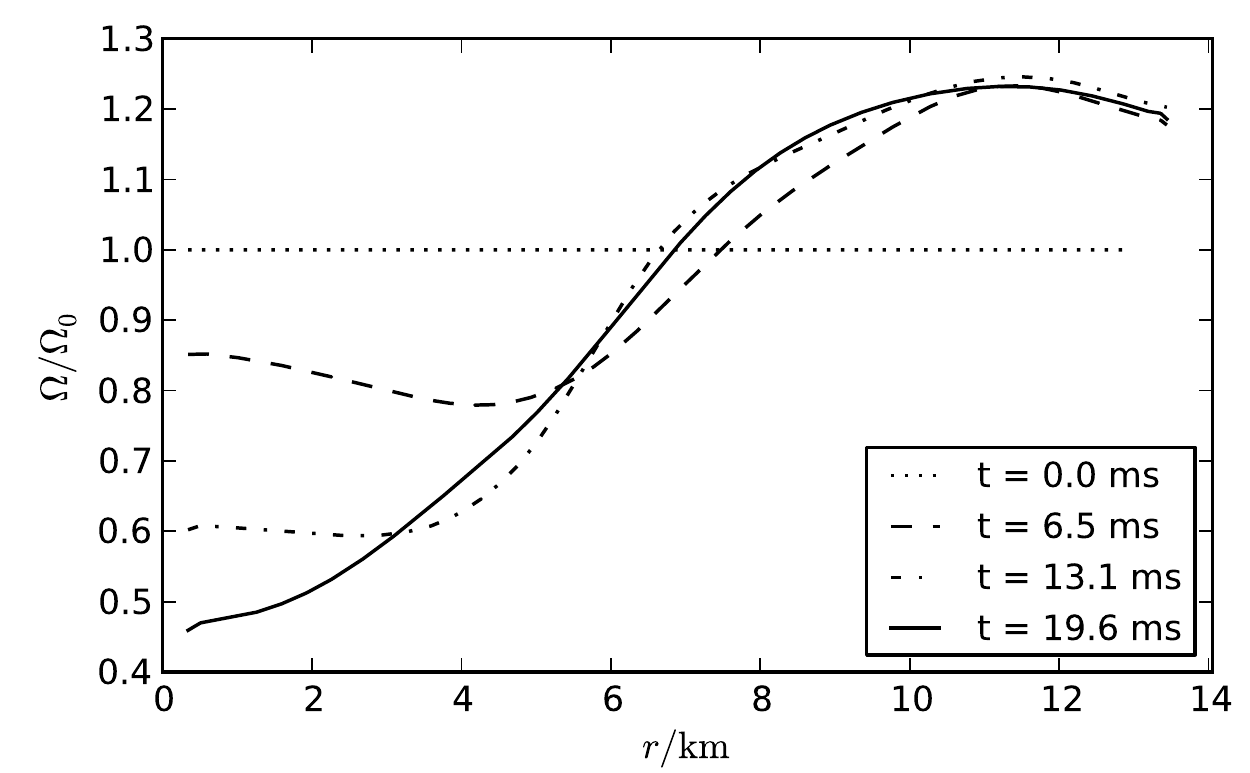}
\caption{\label{fig_bu3_diffrot_eq}
Snapshots at different times 
of the $\phi$-averaged angular velocity in the equatorial plane
versus coordinate radius, for the same setup as \Fig{fig_bu3_diffrot}.
The lines end where the $\phi$-averaged density falls below
0.005 times the central one.
}
\end{figure}

The profile shown in \Fig{fig_bu3_diffrot} is similar both 
in shape and magnitude to the differential rotation
found by~\cite{Tohline2002} for a Newtonian star.
A cut in the equatorial plane is shown in \Fig{fig_bu3_diffrot_eq}.
Near the axis, the rotation rate is slowed down by a factor of two,
while the equatorial rate is increased by a factor $1.2$.
This also agrees well with the profile shown in~\cite{Gressman2002} for the Newtonian case.
It differs however strongly from the profile shown in~\cite{Lin2006}.
On the other hand, the latter was not a $\phi$-average but a cut along the $x$-axis.
The magnitude of differential rotation is large enough 
to cause to a visible deformation of the stellar surface,
as shown in \Fig{fig_bu3_diffrot}.

For the unperturbed models, our code is able to conserve the rotation profile
with errors several orders of magnitude smaller 
than the observed differential rotation. 
The errors in the presence of differential rotation might be larger,
because there is shear motion and because the fluid 
is not co-rotating with the coordinates anymore, particularly near the stellar surface. 
In general, conserving the angular momentum is problematic
with codes based on Cartesian grids.
Therefore we monitor the conserved angular
momentum $J$ defined by \Eref{eq_def_cons}.
For model \BUthree\ with initial amplitude $\Amp=3$, 
the angular momentum is decreasing more or less linearly,
and the total loss at $t=20\usk\milli\second$
is $\Delta J / J = 0.0032$.
At the same time,  
we observe an average angular momentum change $\bar{\Delta} J / J = 0.26$,
much more than the total violation of angular momentum.
It is also greater than the total and average change of angular momentum
introduced by the initial perturbation (caused by second order terms
and surface effects), which are of the order
$\Delta J / J = 0.018$ and $\bar{\Delta} J/J=0.025$.

\begin{figure}
\includegraphics[width=\columnwidth]{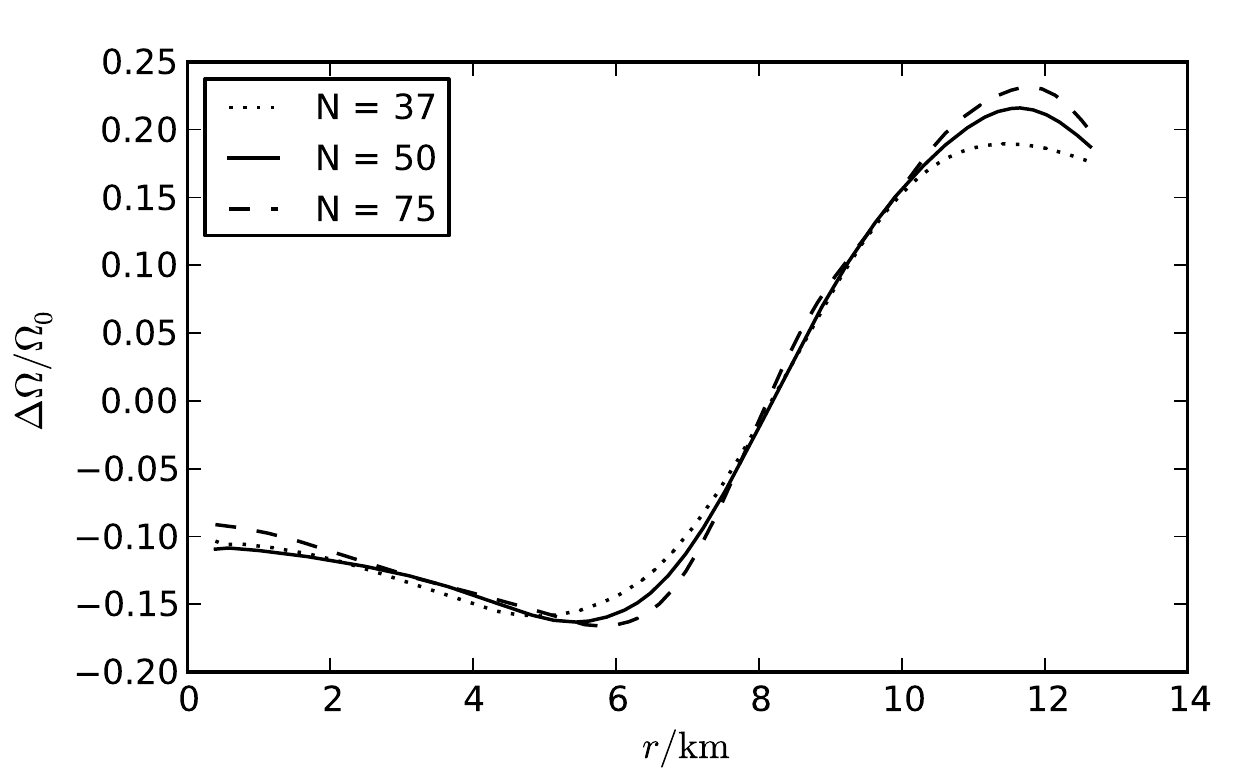}
\caption{\label{fig_conv_diffrot}
Convergence of differential rotation.
Shown is the $\phi$-averaged change of angular velocity after $5\usk\milli\second$
for an $r$ mode of model \BUthree\ with initial amplitude $\Amp=3$,
at different grid resolutions.
}
\end{figure}

Although our computational resources did not permit a full convergence test,
we evolved the first $5\usk\milli\second$ with resolutions 
$N=37,50,75$ points per stellar radius.
For each we sampled the perturbation of $\phi$-averaged angular velocity 
at the end of the simulation along the equatorial plane, 
as shown in \Fig{fig_conv_diffrot}.
On average, low resolution seems to damp the differential rotation
and not to cause it.
To quantify the errors, we computed the $L^2$-norms of the residuals 
and estimated the convergence order $p\approx 1.3$,
which at resolution $N=50$ implies an average error of $\Delta \Omega$ 
around $10 \%$ of the maximum value.
We note that the numerical evolution scheme is second order accurate,
but the treatment of the surface is not. 
From \Fig{fig_conv_diffrot}, one can see that the largest error of $\Omega$ is indeed 
found near the surface.

We conclude that the differential rotation is caused by a redistribution
of angular momentum, and not by numerical errors or contributions already
present in the perturbed initial data.

\subsection{R-mode decay}
Simultaneously with the development of differential rotation,
high amplitude $r$ modes exhibit a rapid decay in amplitude, 
as will be shown in the following.

\begin{figure}
\includegraphics[width=\columnwidth]{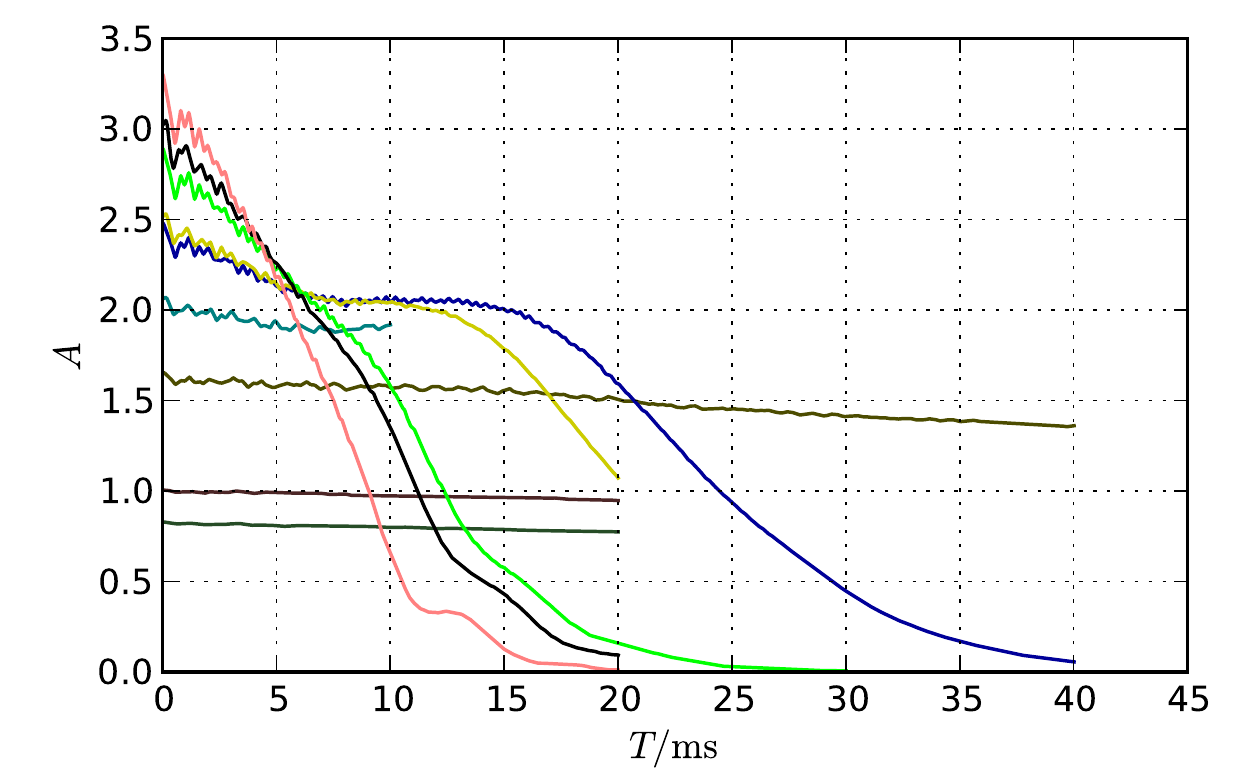}
\caption{\label{fig_bu3_decay}
Decay of the $r$-mode amplitude $\Amp$ defined by \Eref{eq_def_diml_ampl} 
for model \BUthree\ perturbed with different initial amplitudes.
The two curves with initial amplitudes
of $\Amp=2.48$ and $\Amp=2.51$ highlight the steep dependence of the decay time
on the initial amplitude.
}
\end{figure}

\begin{figure}
\includegraphics[width=\columnwidth]{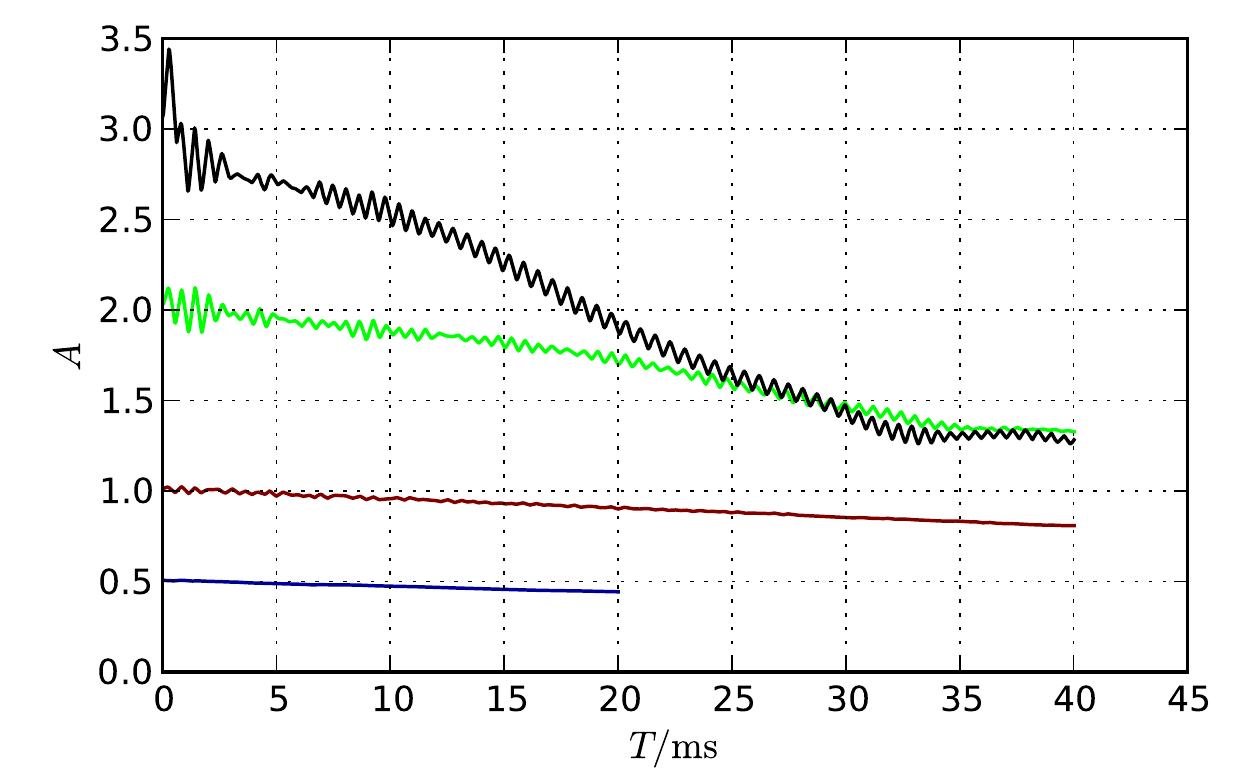}
\caption{\label{fig_bu6_decay}
Decay of the $r$-mode amplitude at different initial amplitudes,
for model \BUsix.
}
\end{figure}

As a measure for the decay, 
we use the dimensionless amplitude $\Amp$ defined by \Eref{eq_def_diml_ampl}.
The results for all our simulations are shown in 
Figs. \ref{fig_bu3_decay} and \ref{fig_bu6_decay}.
All simulations were numerically stable (the shorter ones 
are exploratory simulations).

As one can see, the decay depends crucially on the initial amplitude.
For initial amplitudes of order unity, the decay is comparable 
to the numerical decay, i.e. compatible to no physical decay at all,
while for amplitudes as high as 3, 
the amplitude decreases rapidly.
Interestingly, for some initial amplitudes there is a plateau phase
before the catastrophic decay.
The length of this phase increases rapidly with decreasing initial amplitude.
From our simulations, 
it is unclear whether there is a critical amplitude for the onset
of the decay, or if all $r$ modes decay eventually.
The decay we observe is qualitatively the same as the one
reported in~\cite{Gressman2002}, 
where a Newtonian star is evolved,
exciting an $r$-mode by applying an artificially increased gravitational 
back-reaction that is switched off at given $r$-mode amplitude. 

The fact that the decay rate depends on the \emph{initial} amplitude 
is a strong hint that the main cause
is neither wave breaking nor shock formation, 
since for those the damping strength typically depends on the 
instantaneous amplitude, not on the initial one.
By comparing \Fig{fig_bu3_decay} and \Fig{fig_bu3_diffrot_evol},
we can see that the increase of differential rotation is related
to the $r$-mode amplitude. 
A hypothetical model which would explain 
the accelerating decay and saturating differential rotation is that 
differential rotation causes $r$-mode decay and the presence of the $r$ mode
causes an increase of differential rotation.


A comparison between models \BUthree\ and \BUsix\ shows that 
faster rotation stabilizes high amplitude modes.
For initial amplitudes of order unity, 
doing a comparison is difficult since the damping timescales becomes 
comparable to the numerical damping timescales for both models.

\begin{figure}
\includegraphics[width=\columnwidth]{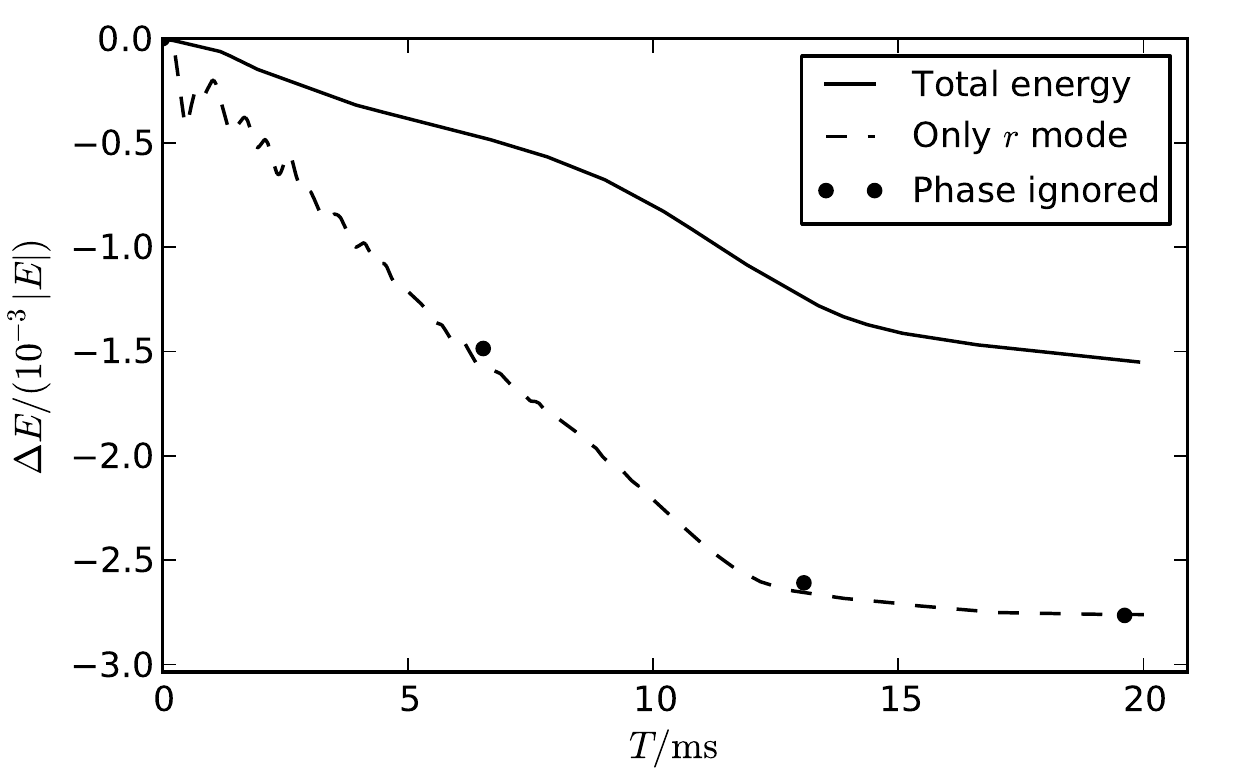}
\caption{\label{fig_cons_energy_bu3}
Energy budget for a decaying $r$ mode of model \BUthree\ with initial amplitude
$\Amp=3$. Shown is the loss of total conserved energy $E$ defined by \Eref{eq_def_cons},
as well as the loss of energy in the $r$ mode, computed by estimating the amplitude 
from the current multipole moment $J_{22}$ and the mode energy from \Eref{eq_def_mode_energy}.
For the points labeled ``Phase ignored'', we took the absolute value of the complex integrand
when computing $J_{22}$.
}
\end{figure}

To answer the question whether the energy of the $r$ mode could
be dissipated directly, e.g.\ via shock formation or numerical dissipation,
we now study the energy budget of the decay,
using the tools from \Sec{sec_cons} and \Sec{sec_mode_energy}.
\Fig{fig_cons_energy_bu3} shows the total conserved energy
as well as an estimate for the energy in the $r$ mode.
The violation of energy conservation is obviously significant,
but it is not sufficient to explain the amplitude decay,
even if all the lost energy is taken from the $r$ mode.
We assume that the mode energy is at least partially transferred to
the differential rotation and deformation of the star described in \Sec{sec_diffrot}. 

\begin{figure}
\includegraphics[width=\columnwidth]{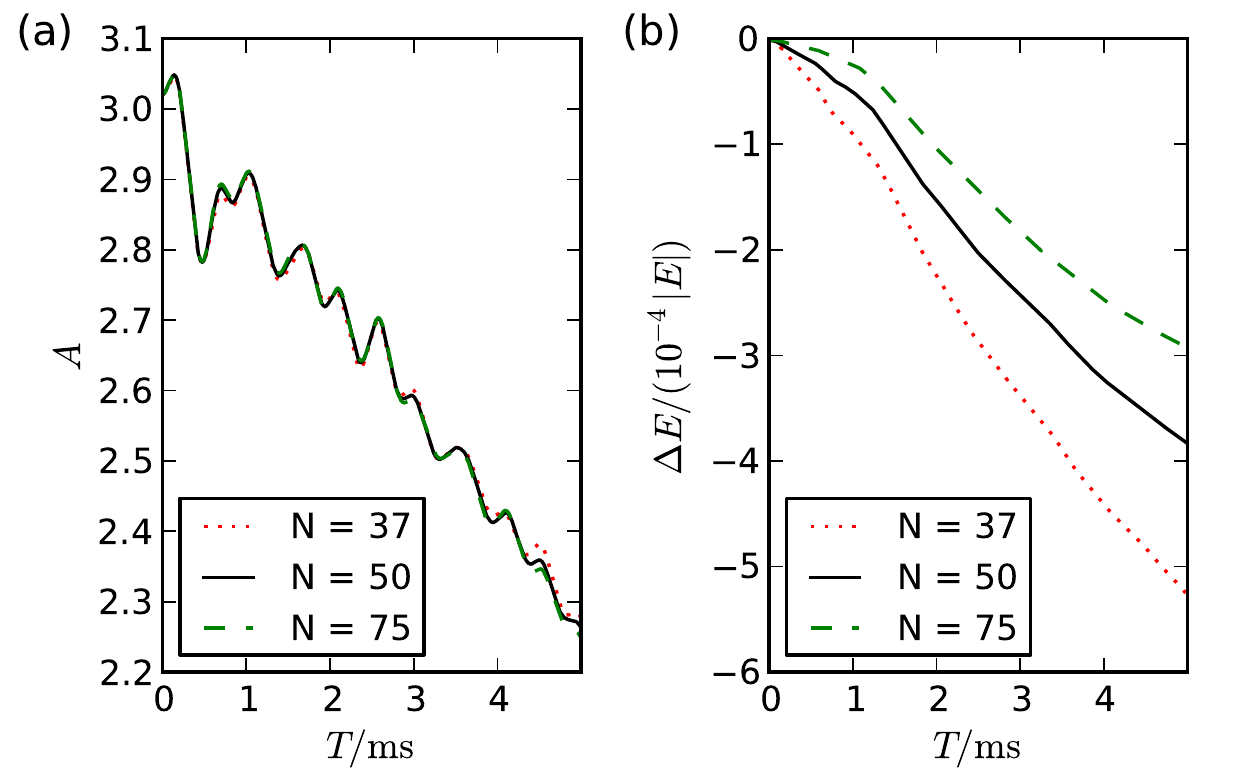}
\caption{\label{fig_conv_test}
Convergence of $r$-mode amplitude (a) and conserved energy (b),
for model \BUthree\ perturbed with an $r$ mode of initial amplitude $\Amp = 3$.
}
\end{figure}

The only effects causing violation of energy conservation 
are formation of shocks in conjunction with the cold EOS,
surface effects like wave breaking,
and numerical dissipation.
To estimate the latter, 
we compare simulations of the first $5\usk\milli\second$ 
at resolutions 37, 50, and 75 points per stellar radius.
As can be seen in \Fig{fig_conv_test}, 
the violation of energy conservation is obviously caused 
for the most part by numerical errors,
although some residual physical effect cannot be ruled out.
However, the comparison also shows that the decay of the $r$ mode 
is computed quite accurately.
This implies that the loss of total energy, which does depend on resolution, 
is unrelated to the loss of $r$-mode energy.

Still looking for the cause of the energy loss,
we produced a movie of the first $5\usk\milli\second$ 
showing density and velocity perturbations in the coordinate planes.
We did not notice any shock formation.
The velocity field is dominated by the $r$ mode.
Right from the start, 
the density field is an overlay of many oscillation modes
including the $r$ mode and the quasi-radial mode.
The presence of other modes is not surprising, 
on the contrary it would be strange if the simple linear scaling of the eigenfunction 
used to excite high amplitude $r$ modes would not excite other modes as well. 
Further, the density perturbation of the $r$ mode itself is weaker than for 
pressure modes of the same kinetic energy.
We noticed an $m=4$ deformation which near the surface becomes 
quite nonlinear, but not enough to cause wave breaking.
However, the resulting numerical dissipation might 
explain part of the energy loss.
In~\cite{Tohline2002}, wave-breaking of the $r$ mode itself 
was observed for a simulation
of a Newtonian star after the $r$-mode amplitude peaked at $\Amp=3.35$.
This does not contradict our results. 
Our maximum amplitude is $\Amp=3.3$, 
and the amplitudes are not directly comparable 
due to the different models and a slightly different definition of the amplitude.

In conclusion, neither numerical errors nor shock formation 
can be the reason for the decay of the $r$ mode, 
and its energy has to be transferred elsewhere,
in particular into differential rotation.
Wave-breaking does not occur in our models, 
but might play a role for even larger amplitudes.

\begin{figure}
\includegraphics[width=\columnwidth]{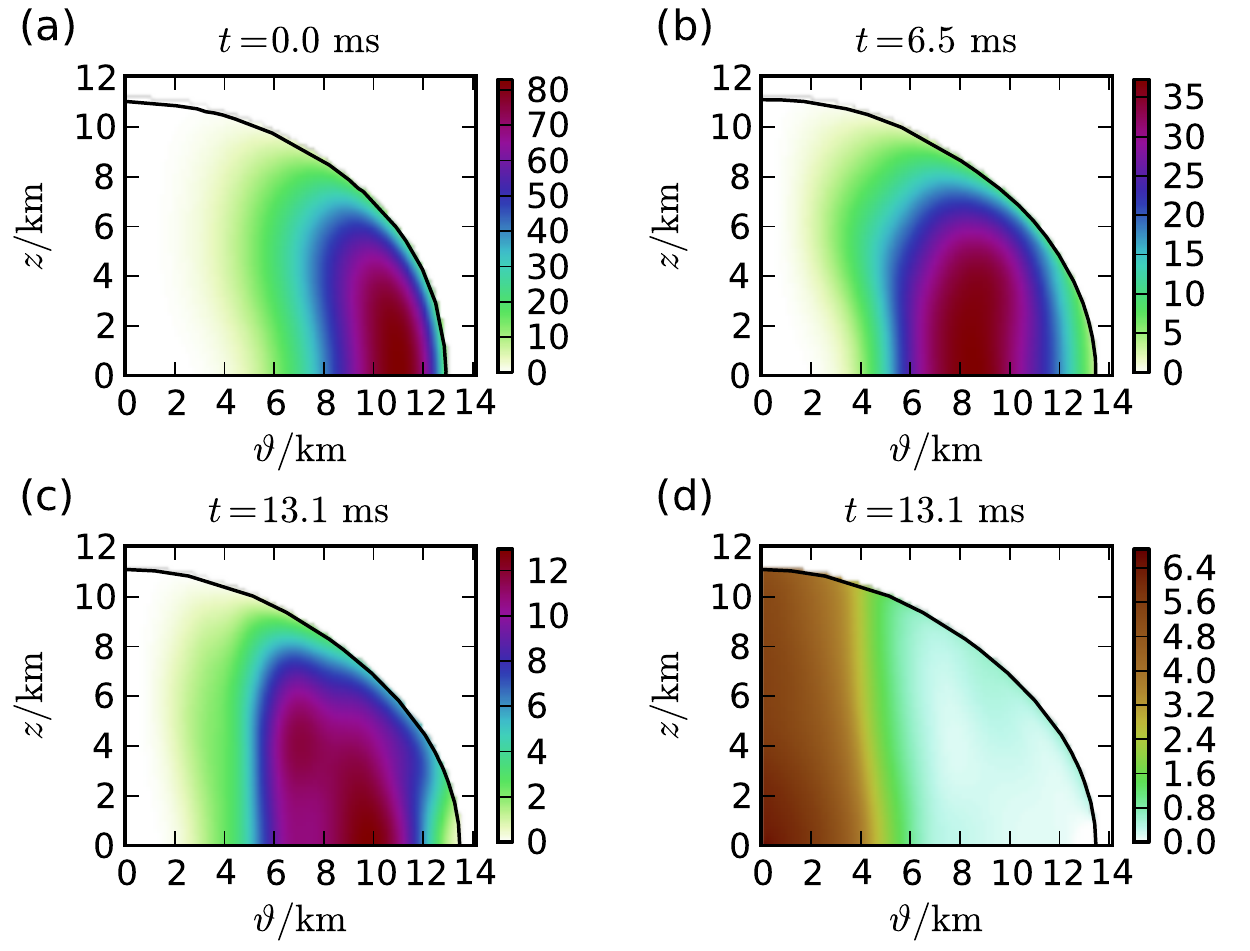}
\caption{\label{fig_bu3_mp_decay}
Time development of the  $\phi$-integrated $l=m=2$ current multipole integrand 
for a nonlinearly decaying $r$ mode of model \BUthree, with initial amplitude $\Amp=3$.
(a)-(d) Absolute value at different times. (d) Complex phase difference to some reference point 
near the end of the decay.
The phase was computed such that the usual phase-jump from $-\pi$ to $\pi$
is avoided, keeping the phase continuous. 
}
\end{figure}

Next, we try to get a more local picture of the $r$-mode decay.
For this we study the integrand of the $l=m=2$
current multipole.
However, we first have to eliminate the influence of differential
rotation, 
because locally the current multipole integrand has a significant
$\phi$-component and therefore couples strongly to differential rotation
(and rotation as such when the background model is not subtracted),
although this cancels out after integration.
It is therefore difficult to interpret plots based on the integrand itself, 
as done in~\cite{Gressman2002}.
Instead, we integrate along the $\phi$-direction to get rid of
any contribution from perturbations with angular dependency $m\neq 2$.
Snapshots at different times are shown in \Fig{fig_bu3_mp_decay}.
While the total amplitude decays, 
the pattern is deformed,
but not destroyed.
This seems to contradict~\cite{Gressman2002}, 
where a breakdown of the mode pattern is reported.
However, it is not obvious to what degree Fig. 5 in~\cite{Gressman2002} is determined by 
the developing differential
rotation instead of the $r$ mode.

We also found that the complex phase, 
which is spatially constant in the linear regime, 
develops a significant variance during the decay.
\Fig{fig_bu3_mp_decay} shows the variance at a late stage.
This seems to imply that parts of the star oscillate out of phase.
Note however that the local amplitude goes to zero at the axis 
roughly quadratically, 
and the phase varies most strongly near the axis where it is
very sensitive to contributions from secondary $m=2$ modes.
We cannot rule out the possibility that the variance is an artifact caused
by other perturbations.
It is worth mentioning that a large phase variance is reported in~\cite{Tohline2002}
as well, although this was under the influence of an artificially large
gravitational back-reaction force driving the $r$ mode.

In order to determine the importance of the phase variance for
our estimate of the $r$-mode energy from the multipole moment,
we recomputed the total multipole moment, 
but used the absolute value of the complex integrand.
The rationale is that the energy, in contrast to the multipole moment,
is not sensitive to an axisymmetric phase shift.
As shown in \Fig{fig_cons_energy_bu3}, the differences are negligible.

We note that the phase variance cannot be attributed to high
amplitudes, since it is still present when the amplitude
has already decayed to values in the linear regime. 
If we assume for a moment that the phase variance of the current multipole
really means that the oscillation of the velocity field 
is spatially out of phase, it follows that some other perturbation
must be present, somehow influencing the $r$ mode.
It is worth noting that the differential rotation profile present after 
the decay phase has the same structure as the phase variance.

\subsection{Search for mode coupling}
\begin{figure}
\includegraphics[width=\columnwidth]{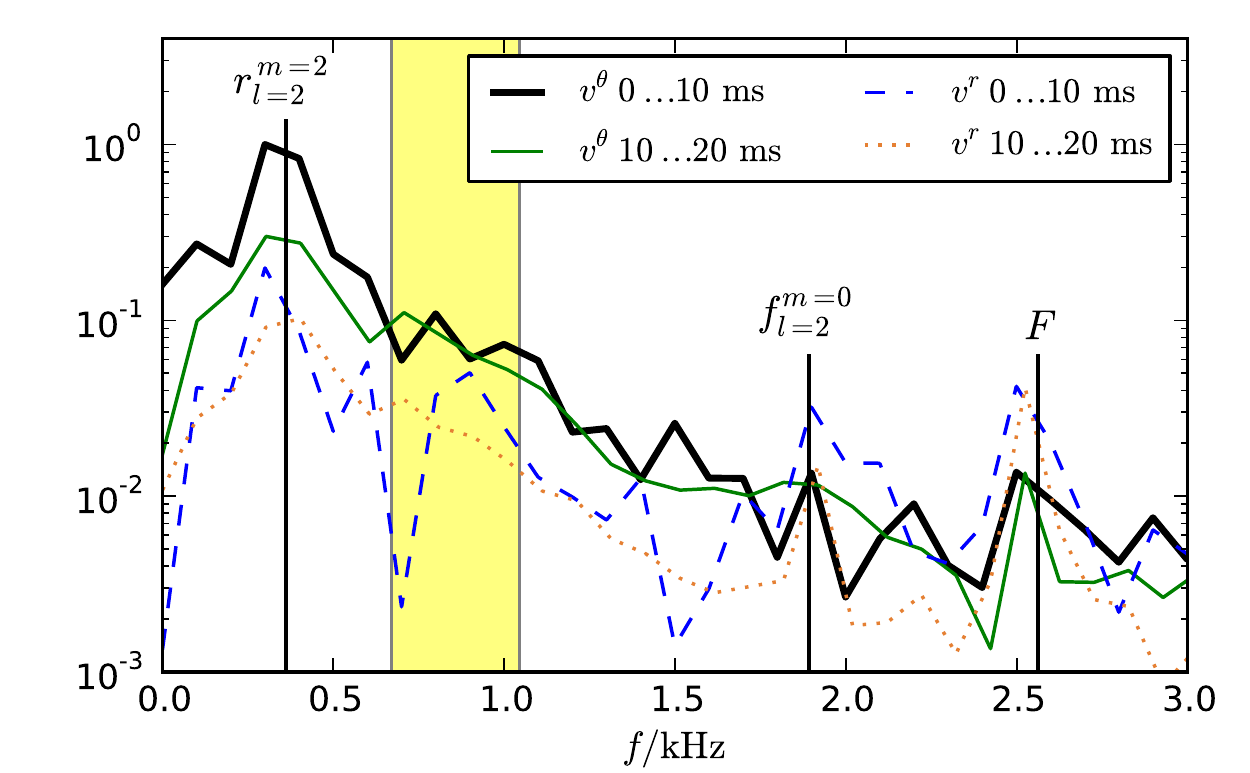}
\caption{\label{fig_spec_evol}
Fourier spectra of the time evolution of $v^\theta, v^r$ at a sample point $x = y =z = 0.24\, R$
of model \BUthree, during the $r$-mode decay with initial amplitude $\Amp=3$.
Shown are two spectra corresponding to the first and second half of the evolution.
For comparison, vertical lines mark the known frequencies of the $r$ mode, the quasi-radial $F$ mode,
and the axisymmetric $f$ mode.
The shaded area marks the transitional band 
between inertial and pressure modes (see main text).
}
\end{figure}

We now discuss possible nonlinear interactions of the $r$ mode
with secondary modes.
For this, we looked at the Fourier spectra of the time evolution at various sample points.
To study the time evolution of the spectra, 
we computed separate spectra for the first and second half of the evolution.
\Fig{fig_spec_evol} shows the spectra for the velocity components $v^r, v^\theta$,
for an initial amplitude $\Amp=3$.
As one can see, the $r$ mode is the dominant contribution.
However, there are significant peaks corresponding to oscillation modes in the inertial
mode spectrum.
We cannot identify those modes since
the frequency resolution of our spectra is insufficient to distinguish
modes in the dense inertial mode spectrum.
Interestingly, there are also peaks in the frequency range
where the 2nd order scalar partial differential equation describing mode oscillations
is of mixed elliptic-hyperbolic type, as discussed in~\cite{Kastaun08}.
It is unclear what the structure of solutions in this range would be, 
and if such solutions exist at all.
It is however possible that the developing differential rotation shifts the 
location of this band.
In any case, the only peaks we identified beside the $r$ mode are the quasi-radial $F$ mode
and the axisymmetric $f$-mode, both quite insignificant for the velocity field.
In the density, they are more visible, since the density perturbation of $r$ modes
in relation to the velocity perturbation is smaller than for pressure modes.

None of the secondary modes with significant amplitude seem to grow.
Only the increasing differential rotation is clearly visible in the spectra of $v^\phi$ (not plotted).
From \Fig{fig_spec_evol}, 
we cannot confirm any mode coupling effect, at least not of the magnitude found
in~\cite{Gressman2002, Lin2006}.
Note however that the spectra in~\cite{Gressman2002, Lin2006} are for an initial amplitude $\Amp=1.6$,
where the decay is slower.
Looking at spectra from our simulations at lower amplitudes,
we sometimes see growing peaks, 
but their amplitudes are tiny compared to the $r$ mode.
There are several different explanations for the discrepancies. 
First, the mode coupling for our models might saturate already during the timespan
covered by the early stage Fourier spectra, 
thus being indistinguishable from secondary oscillations 
excited by the initial perturbation.
Second, the mode coupling reported in~\cite{Gressman2002, Lin2006} might not be the only 
cause of $r$-mode decay, 
and just happens to be less prominent for our models.
Also, we might have chosen the wrong sample points, 
where an important secondary mode happens to be small (Although we also studied some
global quantities like multipole moments).
Our results thus cannot completely rule out mode coupling as the cause
of the $r$-mode decay.

\section{Summary and Discussion}
This work provides new evidence that $r$ modes (with $l=m=2$) 
of uniformly rotating neutron stars with a barotropic EOS 
decay rapidly if their dimensionless amplitude exceeds a model dependent value
of order unity.
The speed of decay depends only on the initial amplitude, 
and the decay does not stop even when the amplitude becomes 
small compared to unity.
The $r$ mode decays more slowly for higher rotation rates (at a fixed central density).
Together with the decay of the $r$ mode, strong differential rotation develops.
The final rotation profile depends roughly on the distance to the axis.
Close to the axis, the rotation is slowed down 
by a factor 0.5,
while near the equator we observed speedups around 1.2.

Our results are the first ones obtained using the relativistic Cowling approximation,
and are in good agreement with previous studies~\cite{Tohline2002, Gressman2002} 
of uniformly rotating stars
which have been treated in the Newtonian framework, 
but without artificially fixing the gravitational field.
Thus, the cause is unrelated both to relativistic effects and to the changes
of the gravitational field induced by the fluid.
Those studies also used a different way to excite the $r$ mode.
While they used an artificially increased gravitational radiation reaction force 
to slowly drive the $r$ mode to large amplitudes,
we perturbed the initial data with linearly scaled exact eigenfunctions.
This is noteworthy because it implies that the decay and the differential rotation
are not sensitive to the amount and composition of other modes in the initial data.

However, we also found some differences. 
The aforementioned Newtonian studies reported the occurrence of 
either wave breaking or strong mode coupling
together with the decay.
We found no wave-breaking, 
although the oscillations near the surface are definitely in the nonlinear regime.
This is not a contradiction. 
Given the different models, our maximum amplitude, 
although comparable to the the reported wave-breaking case, was 
probably just not large enough.
Nevertheless, wave-breaking is not necessary for the $r$-mode decay.
We also cannot confirm the presence of significant mode coupling.
However, since we only analyzed the time evolution of selected sample points
and a few multipole moments, we cannot rule it out completely.

We also studied the energy budget of the process.
For this, we derived a measure for the energy of fluid modes,
which estimates the energy difference between a state perturbed
with the linear eigenfunction
and a ground state with the same angular momentum profile and total mass. 
For this we took advantage of the fact that using
an artificially fixed axisymmetric spacetime
implies the existence of a conserved energy and angular momentum
besides the conserved mass.
During the nonlinear decay, we observed a significant loss of conserved energy
caused mostly, if not completely, by numerical errors,
as we found from convergence tests.
The energy loss of the $r$ mode however was greater, 
and not sensitive to the numerical resolution.
We can thus conclude that the energy of the $r$ mode is not dissipated directly,
e.g.\ due to wave breaking, shock formation, or numerical errors.
Instead it is converted mostly into differential rotation,
which also causes a deformation of the star, 
increasing the equatorial radius by a few percent.

Last but not least, 
we provided eigenfunctions and frequencies of $r$ modes in the relativistic 
Cowling approximation for rapidly rotating stars.
We found excellent agreement with the frequencies found in \cite{Gaertig2010}, 
thus validating those results.
Within numerical accuracy, the eigenfunctions are smooth.
Eigenfunctions and frequencies are still very similar to the 
Newtonian slow rotation case.
We also estimated the gravitational luminosity, wave strain, and
angular momentum loss caused by $r$ modes, 
and found that the current multipole is still strongly dominant
for the rapidly rotating case.
For our models, 
an $r$ mode with unit dimensionless amplitude at a distance of $10\usk\mega\parsec$
causes a wave strain of the order $10^{-24}$.

Finally we would like to speculate a bit.
Although the decay timescale diverges quickly with decreasing amplitude,
we have no proof that the effect vanishes completely below some critical
amplitude. 
Let us assume that the $r$-mode growth due to the CFS instability 
and the decay described in this work are balanced 
at some amplitude too small to be relevant as a source for detectable gravitational radiation.
It is plausible to assume that the $r$-mode energy loss is still converted
to differential rotation. 
The effect would then be cumulative as long as the CFS instability is active.
It might well be that the CFS instability of the $r$ mode does not induce
large amplitude oscillations, but differential rotation of similar energy.
This effect, which is of course purely hypothetical, 
might be important for the time evolution of 
the rotation profile of newborn neutron stars, 
and also may lead to strong amplification of the magnetic field.

\begin{acknowledgments}
Our numerical computations have been performed using the \textsc{hpc-bw} cluster of
the University of T\"ubingen, the \textsc{hg1} cluster at SISSA, 
and the Italian \textsc{cineca} cluster. 
We would like to thank Kostas Kokkotas, John Miller, 
and Luciano Rezzolla for useful suggestions concerning the manuscript.
\end{acknowledgments}

\bibliography{article}
\bibliographystyle{apsrev}

\end{document}